# Structured illumination multimodal 3D-resolved quantitative phase and fluorescence sub-diffraction microscopy

Shwetadwip Chowdhury[1], Will J. Eldridge[1], Adam Wax[1], and Joseph A. Izatt[1,*]

**Authors information:**
[1]Duke University, Biomedical Engineering Department
1427 FCIEMAS, 101 Science Drive, Box 90281
Durham NC, 27708

**\*Corresponding Author:** joseph.izatt@duke.edu

## Abstract

Sub-diffraction resolution imaging has played a pivotal role in biological research by visualizing key, but previously unresolvable, sub-cellular structures. Unfortunately, applications of far-field sub-diffraction resolution are currently divided between fluorescent and coherent-diffraction regimes, and a multimodal sub-diffraction technique that bridges this gap has not yet been demonstrated. Here we report that structured illumination (SI) allows multimodal sub-diffraction imaging of both coherent quantitative-phase (QP) and fluorescence. Due to SI's conventionally fluorescent applications, we first demonstrated the principle of SI-enabled three-dimensional (3D) QP sub-diffraction imaging with calibration microspheres. Image analysis confirmed enhanced lateral and axial resolutions over diffraction-limited QP imaging, and established striking parallels between coherent SI and conventional optical diffraction tomography. We next introduce an optical system utilizing SI to achieve 3D sub-diffraction, multimodal QP/fluorescent visualization of A549 biological cells fluorescently tagged for F-actin. Our results suggest that SI has unique utility in studying biological phenomena with significant molecular, biophysical, and biochemical components.

# I. Introduction

Optical microscopy has played a crucial role in advancing frontiers of the biological sciences by allowing high-resolution, non-invasive visualization of important biological samples. Although developments and advances in optical design and manufacturing have made available high-resolution objectives with unprecedented numerical aperture (NA), microscopy faces a fundamental physical diffraction limit that can preclude visualization of important sub-cellular features in biological samples[1,2]. In response, several imaging techniques have been developed which allow far-field sub-diffraction resolution imaging using a variety of unique and innovative mechanisms[3,4].

Sub-diffraction imaging techniques introduced thus far operate in two main regimes: 1) imaging via spatially-coherent diffraction, or 2) imaging via spatially-incoherent fluorescence. Synthetic aperture (SA) is a popular choice for imaging in the first regime, and operates by using oblique illuminations to spatiotemporally encode a wider frequency support into the final image than directly allowed by the microscope's physical aperture[5-7]. Applications of SA have resulted in both high-resolution imaging, where (Sparrow) resolutions of <100 nm have been achieved[8], and high-throughput imaging, where gigapixel-scale images with resolutions > 5x over the diffraction limit have been obtained[9,10].

Sub-diffraction techniques for imaging in the fluorescent regime, often referred to as "super-resolution" techniques, typically require the sample's fluorescent labels to have either photoswitching or depletion capabilities. Photoactivated localization microscopy (PALM) is a prominent example that uses photoswitchable fluorophores to localize individual emitters at sub-diffraction resolutions per acquisition, before combining them into a final super-resolved image[11]. Stimulated emission depletion (STED) is another successful example that utilizes point-scanning to directly minimize the size of the scanned focal spot by saturated stimulated emission with two synchronized ultrafast laser sources[12]. Such techniques have found tremendous success in biological imaging and have achieved resolutions well below 100 nm.

These two regimes operate on fundamentally different mechanisms, and thus enable distinct and complementary biological observations. Fluorescence imaging is the standard for molecular-specific, background-free, cellular imaging, and has enabled great insights into gene expression, protein interaction, cytoskeletal organization, endocytic dynamics, organelle structures, intracellular transport, cytokinesis, and general intracellular dynamics [13-18]. Coherent diffraction imaging is the choice technique to image unstained and minimally prepared cellular samples with endogenous contrast to extract quantitative and biologically relevant parameters. Examples of coherent-diffraction imaging include quantitative-phase (QP) and light-scattering imaging, which can noninvasively probe structural, mechanical, biophysical, and biochemical properties of cells, and have been used for analysis of whole-cell morphology, mass, shear stiffness, refractive-index/optical-path-length (OPL), dispersion spectroscopy, and absorption/scattering[19-23]. Important biomedical applications even include non-invasive and quantitative measurements of hemoglobin oxygenation saturation[24] and detection of different cancers types[25]. In both fluorescent and coherent-diffraction regimes, biological insight is enhanced with sub-diffraction resolution – unfortunately, because the two regimes are fundamentally different, a sub-diffraction resolution method applicable to both has been difficult to find. This poses an obstacle in microscopy, as users conventionally must choose between either unimodal sub-diffraction or multimodal diffraction-limited imaging. Such a choice can prevent a synergistic, multimodal analysis of individual sub-cellular components beyond the diffraction limit and can hinder a cohesive understanding of biological morphology and function.

In this work, we demonstrate that structured illumination (SI) microscopy is a sub-diffraction technique compatible with both coherent-diffraction and fluorescent imaging. Indeed, SI is already an established technique which is conventionally associated with fluorescent super-resolution and has become popular due to its speed, cost, and simplicity[13-15]. However, SI also has applications in coherent imaging and has shown promise for sub-diffraction quantitative-phase (QP) imaging of diffractive samples[26-28]. Because SI enables fluorescent and coherent-diffraction sub-diffraction imaging, it offers a unique capability for biologists to conduct multimodal studies that probe relationships between the biophysical/biochemical properties of a cell with its molecular properties/processes. In the following sections, we experimentally demonstrate, for the first time in our knowledge, a single optical system that uses SI to generate 3D QP and fluorescence visualizations at sub-diffraction resolutions, with potential applications for future multimodal analysis of sub-diffraction cellular components.

# II. Theoretical framework

## A. Imaging transfer functions.

Diffraction theory describes the maximum lateral spatial frequencies $k_{\parallel,F}$ and $k_{\parallel,QPM}$ and maximum axial spatial frequencies $k_{\perp,F}$ and $k_{\perp,QPM}$, respectively, that can be observed through a microscope for both conventional fluorescence and QP microscopies. For the case where QP and fluorescent modalities share the same physical microscope and the light used for QP imaging serves the dual purpose of also being the excitation for fluorescent imaging, these maximum spatial frequencies are given by $k_{\parallel,F} = 2\text{NA}/\lambda_{em}$, $k_{\parallel,QP} = \text{NA}/\lambda_{ex}$, $k_{\perp,F} = n(1 - \cos\theta)/\lambda_{em}$, and $k_{\perp,QP} = n(1 - \cos\theta)/\lambda_{ex}$[29]. Here, $\lambda_{ex}$ and $\lambda_{em}$ are the excitation and emission wavelengths used for QP and fluorescent imaging, respectively, $n$ is the index of refraction of the medium, $\theta$ is the maximum half-angle of light that the detection objective supports, and $\text{NA} = n\sin\theta$ is the detection objective's numerical aperture. In three-dimensional Fourier space, these spatial frequencies set the bounds of observable spatial frequency content that can be passed through the microscope (note that the spatial frequency bounds for QP and fluorescence are only applicable in the electric-field and optical intensity regimes, respectively, so direct comparisons are inappropriate). In accordance with diffraction theory, these regions of observable spatial frequencies define the systems' transfer functions (TF) and take the shape of a spherical cap, mathematically a subsection of Ewald's sphere, in QP imaging and a torus-like structure in fluorescence imaging (Fig. 1)[2,30]. Note that $k_{\perp,QPM}$, does *not* equate to QP imaging's axial resolution – because QP imaging's TF has infinitesimal axial frequency support, the sample's diffracted wave vector with the $k_{\perp,QPM}$ axial component propagates through the whole image volume and offers little optical sectioning[31,32].

The spatial frequencies outside the QP and fluorescent TFs are typically unobservable and are the target for visualization by sub-diffraction resolution imaging. SI achieves this by aliasing information into the TF by illuminating the sample with a spatially modulated illumination pattern. Because this aliasing effect happens naturally when two spatial patterns overlap regardless of fluorescent or diffractive imaging, SI can serve as a generalized platform for multimodal sub-diffraction imaging, up to a resolution gain factor of 2. Nonlinear SI, which uses fluorescent nonlinearities to achieve resolution gains of >2, is not considered here[33]. In this work, we demonstrate that SI allows 3D sub-diffraction multimodal QP and fluorescent imaging. We refer the reader to the work by Gustafsson *et al.*[30], which beautifully illustrates the concept behind SI for 3D fluorescent imaging – in the next section, we introduce an analogous approach for QP imaging.

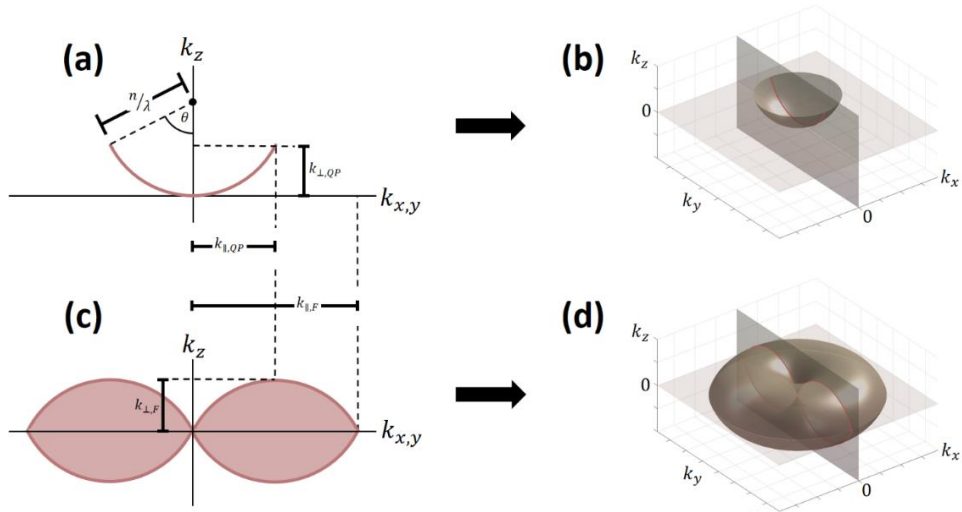

**Figure 1 | 3D transfer functions in QP and fluorescence imaging, assuming $\lambda = \lambda_{ex} \approx \lambda_{em}$.** (a,b) Cross-sectional plot and 3D rendering demonstrate that the transfer function in QP imaging forms a spherical cap (commonly referred to as Ewald's Sphere). Dimensional parameters of this cap are determined by wavelength, microscope objective NA, and immersion oil refractive index. Autocorrelation of this cap corresponds to the transfer function for fluorescence imaging. (c,d) Cross-sectional plot and 3D render show this region to have a filled torus-like shape. Note that direct comparisons between (a) and (c) are inappropriate because spatial frequencies in QP and fluorescence imaging are measured in the electric-field and intensity regimes.

## B. Principle of SI-enabled 3D QP visualization.

The SI framework for sub-diffraction resolution imaging is typically modelled mathematically as a modulation of the spatial frequencies in the sample by those in an illumination pattern, which is then imaged through a system that operates under constraints of linearity and translation-invariance. The illumination's modulation of the sample's spatial frequencies results in resolvable "beat" frequencies (i.e., Moiré patterns) that allow reconstruction of sample spatial frequencies that fall outside the system's diffraction limit[34]. This mathematical treatment does not make any fundamental assumptions on coherence, and is thus equally applicable to both fluorescent and coherent-diffraction imaging (as long as complex electric-field is imaged in the coherent imaging case so that the requirement for linear and translation-invariant imaging is satisfied). Indeed, SI's ability to enable sub-diffraction resolution coherent QP-microscopy (SI-QPM) in two dimensions has been clearly demonstrated[26-28].

To intuitively extend the conceptual framework towards 3D coherent imaging, we note that for coherent-diffraction imaging, SI is equivalent to multiplexed oblique illumination, where each tilted plane wave in the illumination's plane-wave-decomposition contributes linearly to the electric-field at the image plane[35]. It follows from Fourier theory that because on-axis plane wave illumination results in a spherical-cap TF centered at the origin of frequency space, tilted plane-wave illumination, mathematically described with a phase ramp, displaces the TF in frequency space by an amount equal to the tilt angle. This concept is identical to and ubiquitously used in optical diffraction tomography (ODT)[22,29,36].

SI-QPM's main difference from conventional ODT is in implementation – because SI linearly multiplexes tilted illuminations onto the sample to achieve its structured pattern, the imaged field is a linear superposition of the sample's spatial frequencies from correspondingly shifted TFs. In the case of a sinusoidal structured pattern, phase shifting the pattern allows for analytical solution for these spatial frequencies, which can then be digitally shifted to their correct regions in 3D frequency space[33,34]. Because the coherent TF is a spherical cap with infinitesimal axial frequency support, simply illuminating with structured patterns with spatial frequencies at the maximum-allowed magnitude, as is typically done to maximize lateral resolution gain in fluorescent SI, leaves much of the axial frequency space uncovered. To thoroughly cover 3D frequency-space, the tilt angle of the illuminations must be incremented. Conventionally, ODT accomplishes this by scanning the tilt angle of the illumination beam directly with a pair of scanning mirrors. SI can achieve the same effect by illuminating with spatial sinusoids of varying spatial frequencies (Fig. 2).

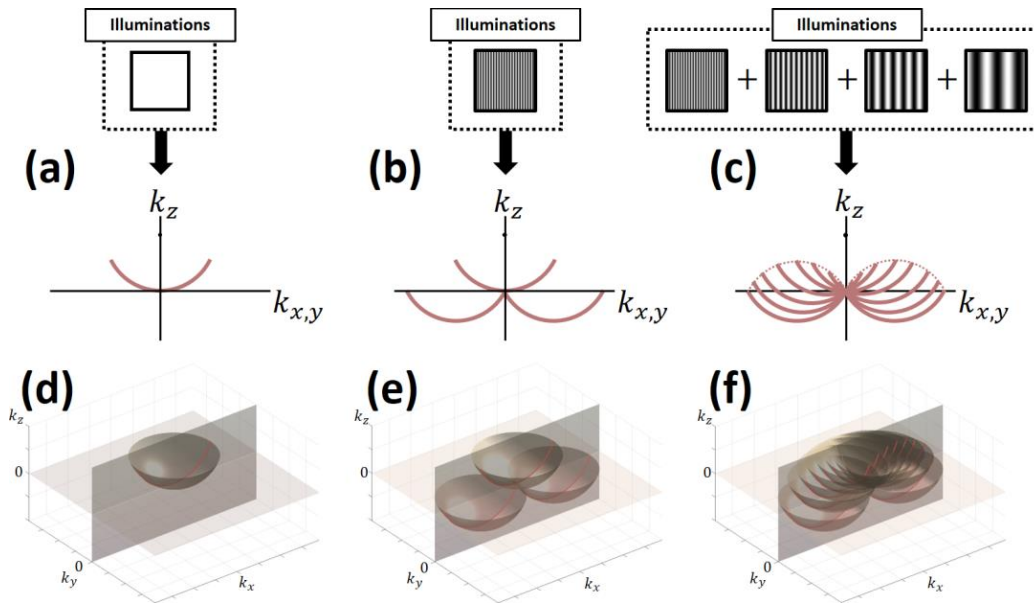

**Figure 2 | Filling out 3D frequency space with structured illumination.** Cross-sectional plots show how axial frequency space is filled with (a) uniform illumination, (b) sinusoidal illumination with maximum allowable spatial frequency, and (c) a summation of sinusoidal illuminations of varying spatial frequencies. Note that illuminating with simply one spatial frequency is not sufficient to fill out axial frequency space in coherent-diffraction imaging, as it is in fluorescent imaging. (d,e,f) Corresponding 3D renderings show 3D frequency space being filled by Ewald caps.

Coherent SI and ODT are conventionally considered separate imaging techniques, and so their connection as intuitively explained above may not be immediately obvious. For the interested reader, Supplementary Note 1 rigorously formulates the theory for SI-enabled 3D QP sub-diffraction imaging upon the fundamental principles of 3D fluorescent SI super-resolution introduced by Gustafsson *et al*[30]. Though this formulation was derived independently of any consideration of ODT's conventional framework, the final conclusions in Supplementary Note 1, summarized in Supplementary Figs. 2(g-j), is mathematically identical to those of ODT.

## C. Reporting diffraction-limited resolution.

Reporting on an imaging system's resolution usually involves reporting some metric of "minimum-resolvable distance". We choose to report theoretical diffraction-limited resolution in terms of the Abbe limit, which gives the cycle-period of the highest spatial frequency of the sample that is transferred by the microscope onto the image plane. In our case of DC-centered, symmetric, and *filled* lateral TFs, diffraction-limited lateral Abbe resolutions are the reciprocal of the spatial frequency bounds described above – QP and fluorescent imaging have Abbe lateral resolutions of $d_{\|,QP} = 1/k_{\|,QP} = \lambda_{ex}/\text{NA}$ and $d_{\|,F} = 1/k_{\|,F} = \lambda_{em}/2\text{NA}$ in the electric-fields and optical intensity domains, respectively. As mentioned previously, due to its infinitesimal axial TF extent, conventional QP imaging has little axial resolution[31,32]. SI-enabled 3D QP, however, has an expected Abbe axial resolution in electric-field of $d_{\perp,QP} = 1/k_{\perp,QP} = \lambda_{ex}/n(1-\cos\theta)$. Similarly, fluorescent imaging has an Abbe axial resolution of $d_{\perp,F} = 1/k_{\perp,F} = \lambda_{em}/n(1-\cos\theta)$ in optical intensity.

We choose the Abbe standard for "minimum-resolvable distance" because of its straightforward applicability to both coherent and incoherent modalities. Other popular alternatives that involve basic parameters of the image's point-spread-function (FWHM, Rayleigh, Sparrow) have complex interpretations in coherent imaging[1,37-40]. In the case of intensity-based coherent imaging, where the imaged intensity is nonlinearly related to sample structure, such interpretations may be outright misleading. Even in the case of electric-field imaging via holographic or computational methods, accurate interpretations must incorporate considerations of the system's coherence properties or the sample's phase dependences. In the case of a fully coherent system and a sample of uniform phase, we show in Supplementary Fig. 3 that the Abbe limit is a conservative metric for resolution and that other reporting schemes for resolution may be significantly more attractive – we emphasize however that such resolution "improvements" are not indicative of an actual difference in the system's PSF or imaging performance.

We also note that in the case of incoherent diffraction-based imaging, such as brightfield, darkfield, phase-contrast, or differential-interference contrast (DIC) microscopies, the lateral Abbe diffraction limit is generally considered to be $d_{\|,inc} = \lambda/(\text{NA}_{illum} + \text{NA})$, where $\lambda$ is the generalized imaging wavelength and $\text{NA}_{illum}$ is the numerical aperture of the illumination objective (i.e., condenser lens) that sets the angular range available for illumination[41,42]. In the case in which the illumination and detection objectives have equal numerical apertures, such that

$NA_{illum} = NA$, the resolution limit $d_{\parallel,inc} = \lambda/2NA$ matches the resolution limit $d_{\parallel,F}$ of fluorescent microscopy. Because typical QP microscopies use on-axis plane wave illumination, it is tempting to conclude that the resolution limit $d_{\parallel,QP} = \lambda/NA$ is simply a special instance of $d_{\parallel,inc}$ under the constraint $NA_{illum} = 0$. One could then naturally argue that it is unfair to consider $\lambda/NA$ as the diffraction limit when the sample could conceivably be illuminated with the full angular range allowed by the illumination objective to directly achieve $\lambda/2NA$ resolution imaging without the addition of SI. To respond to this, we assert that the general expression for $d_{\parallel,inc}$ makes the key assumption of incoherent illumination, and thus is not an appropriate standard for coherent imaging. Indeed, simply illuminating the sample with coherent waves spanning the full range of illumination angles allowed by the illumination objective equates to illuminating the sample with a speckle pattern, which is unsuitable for widefield imaging. Hence, the general standard for coherent widefield imaging in the QP-imaging community is to image the sample with a flat electric-field illumination background, achieved with a single on-axis coherent beam (i.e., $NA_{illum} = 0$), and consider the resulting resolution limit of $d_{\parallel,QP} = \lambda/NA$ as the coherent system's diffraction limit[5,9,32,42,43].

# III. Methods.

## A. Optical System.

The general design of our system is illustrated in Supplementary Fig. 4. As shown, our system uses single-mode broadband light (NKT Photonics, EXW-6) spectrally filtered to 488 ± 12.5 nm (Semrock, FF01-482/25), which serves the dual purpose of being the illumination and excitation for QP and fluorescent imaging, respectively. This light was collimated and passed through a 50:50 polarization beam splitter (PBS, Thorlabs, PBS251) before being incident onto an amplitude spatial light modulator (SLM, Holoeye, HED 6001). Due to the SLM being nematic, and thus capable of accepting non-binary inputs, sinusoidal patterns were programmed into the SLM. In the domain of the SLM's coordinate space, the minimum spatial-period of written patterns was limited to ~27 um, corresponding to approximately 3.4 SLM pixels (SLM pixel pitch = 8.0 um) and thus satisfying the Nyquist limit for SLM pixel sampling. The pattern written onto the spatial light modulator was passed through the first 4f system (L1 → L2) before being imaged through the second 4f system (L3 → OBJ) onto the sample. To ensure faithful generation of a sinusoidal pattern at the sample, extraneous diffraction orders resulting from the SLM's pixilation were spatially filtered out by an adjustable iris diaphragm (F, Thorlabs, ID25) placed in the Fourier plane of the first 4f system. The focal length of L3 was chosen so that the desired ±1 diffraction orders arising from a ~27 um spatial-period sinusoidal pattern would be refocused to points near the opposite edges of the back focal plane of OBJ. Diffraction and fluorescence from the sample are collected in transmission from the sample through a detection objective (matched in NA to the imaging objective) and is magnified by the second 4f system (OBJ → L4). The sample's fluorescence signal is split from the diffraction signal by a dichroic mirror (DM, Thorlabs, DMLP505) and further spectrally filtered (SF, Thorlabs, FEL500) before imaging onto the first camera (CMOS-F, Pixelink). The sample's diffraction signal, after being split from the fluorescence via the DM, was passed through a diffraction-phase setup, where a grating (DG, Edmund Optics Ronchi 60 lpmm) was placed at the conjugate image plane to the sample to split the signal into various diffraction orders. A mask (M) was positioned at the Fourier plane of DG to physically block the -1st diffraction order while completely passing the +1st order. The mask also contained a 20 um pinhole (PH, Edmund Optics, 52-869) to spatially filter the 0th order to generate a uniform wavefront reference beam to interfere with the +1st diffraction order at the second camera (CMOS-QP, Pixelink). Due to broadband common-path off-axis interference, high temporal phase stability and low coherent noise (noise variance of < 0.001 rad for widefield imaging) were achieved[26]. Supplementary Fig. 4(c) shows in more detail how the mask was positioned relative to the diffraction orders arising from DG and the SLM.

For experiments involving calibration microspheres (Fig. 3, Supplementary Figs. 6 and 7), a 60X, 1.0 NA Nikon Physiology objective lens was used. For the experiments involving A549 cells (Fig. 4, Supplementary Figs. 8 and 9), this objective was replaced with a 40X, 1.3 NA Zeiss objective lens. The focal length of the condenser lens L4 was chosen to set the system magnification such that a coherent diffraction limited spot was sampled by ~6x6 camera pixels (6.7um pixel pitch). This oversampling, in conjunction with the off-axis interference frequency set by DG (60 lpmm), was sufficient to extract the sample's complex electric-field via standard digital off-axis holography techniques[42-44]. Examples of raw interferogram acquisitions alongside associated Fourier distributions are shown in Supplementary Fig. 5, to demonstrate clear isolation of sample's complex electric-field information in Fourier space.

## B. Image Acquisition.

Custom acquisition software was written in MATLAB. Three beam interference, with an unblocked 0 order, was used to generate periodic illumination patterns for 3D fluorescent and QP SI. Our sample was axially scanned during image acquisition at increments of 200nm, satisfying the Nyquist requirement set by our QP and fluorescent axial diffraction limits. A total of 60 axial slices were taken for a single volume. There was no physical change in the optical system between 3D fluorescent and QP SI imaging – however, the patterned illumination procedures for the two modalities were different and so the fluorescent/QP acquisitions were not simultaneous.

For 3D fluorescent SI imaging, the maximum allowed spatial frequency was used for the illumination pattern, such that the component ±1 orders were at the edge of the illuminating objective's back focal plane. For each imaged z-plane, acquisitions were taken for five phase-shifts, spaced $2\pi/5$, per rotation of the illumination pattern, with two rotations, spaced $\pi/2$ radians apart. In total, this corresponded to 600 raw acquisitions to reconstruct a single volume. Camera integration time was set to 120 ms per acquisition for sufficient fluorescent SNR.

For 3D QP SI imaging, the spatial frequency magnitude of the periodic illumination pattern was incremented 10 times through the domain $[0, NA/\lambda]$, per imaged z-plane. For each spatial frequency, acquisitions were taken for five phase-shifts, spaced $2\pi/5$, per rotation of the illumination pattern, with six total rotations, spaced $\pi/3$ radians apart. In total, this corresponded to 18000 raw acquisitions to sub-diffraction resolve a single volume. Camera integration time was set to 15 ms per acquisition to average out the high-frequency temporally-fluctuating "flickers" inherent with our SLM device. Supplementary Fig. 5 illustrates examples of raw interferograms and associated Fourier transforms from which complex-valued sample electric-fields are reconstructed.

For both fluorescent and QP imaging, phase-shifting of the spatial illumination pattern allowed solving for the sub-diffraction resolution spatial frequency components via linear system inversion. These components were then translated back to their appropriate regions in frequency space[26,34] before being combined (Supplementary Note 1). Standard Wiener deconvolution methods were applied to account for the non-uniform weighting of the final computed TF. Standard whole-image adjustments (brightness, contrast) were consistently applied across image datasets, in accordance with accepted practices for image presentation[45,46].

## C. Sample Preparation.

Microsphere Phantom Preparation. We prepared calibration samples of 400nm, 520nm, and 770nm diameter polystyrene microspheres (BangLaboratories). In order to attach the beads to a surface, 10 uL of the microsphere dilutions (2 uL stock-solution / 500 uL ethanol) were placed onto #1.5 coverslips and allowed to dry. 1X phosphate buffered solution (PBS) was placed over the regions of interest and served as the appropriate immersion fluid for our 1.0 NA Nikon Physiology objective lens.

Cell Preparation. A549 lung cancer cells were cultured using Dulbecco's Modified Eagle Medium (DMEM) supplemented with 10% fetal bovine serum and 1 µL/mL pen-strep. Cells were plated at low-density onto #1.5 coverslips and allowed to attach to the substrate overnight. Cells were fixed using a 4% paraformaldehyde in PBS. Alexa Fluor 488 phalloidin (Life Technologies) was used to stain filamentous actin following the manufacturer's suggested protocol. Following staining, an adhesive spacer and coverslip was placed on top of the sample to ensure a uniform PBS layer above the imaging field of view. Oil of refractive index n=1.51 was placed over the region of interest and served as the appropriate immersion fluid for our 40X 1.3 NA Zeiss objective lens.

# IV. Results

## A. 3D visualization of microspheres with SI-QPM.

The design of our SI-QPM system largely follows our original SI-DPM system, which achieved QP imaging via common-path off-axis holography[26]. One notable difference in our current SI-QPM system, however, is that a spatial-light-modulator (SLM) is used instead of a physical grating to generate structured patterns at the sample. This is necessary to have pixel-addressable control over the patterns, which in turn allows tuning the patterns' translations, rotations, and spatial-frequency magnitudes without physically moving or replacing physical gratings. Our system uses broadband, single-mode, illumination at $\lambda = 488 \pm 15$ nm with an imaging objective of NA = 1 designed for imaging through immersion media with refractive index $n = 1.33$. This yields a lateral diffraction limit of $d_{\parallel,QP} = \lambda/\text{NA} \approx 488$nm and an axial diffraction limit for tomography of $d_{\perp,QP} = \lambda/[n(1 - \cos\theta)] \approx 1.0$um. Our specific acquisition procedures used to reconstruct a QP tomographic volume via SI are detailed in Methods.

SI's capability to enable 2D lateral QP subdiffraction resolution[26-28] was verified by our updated SI-QPM system with calibration and biological samples (shown in Supplementary Figs. 6(a-d), 8(a-d), and 9(a-d)). Here, however, we highlight the importance of SI-enabled 3D QP by demonstrating how a lack of axial resolution may affect and degrade lateral visualization of structures even within the system's diffraction limit. We imaged a monolayer sample of 520nm polystyrene microspheres ($n = 1.60$ at $\lambda = 488$nm) through an immersion medium of 1x Phosphate Buffered Solution (PBS). Figs. 3(a,b) compare the SI-enhanced and conventional widefield (WF) QP images, respectively, of a central x-y slice through an imaging volume of this sample, demonstrating superior visualization with SI enhancement. Figs. 3(c,d) show depth-slices through the location marked by the dashed yellow line in Figs. 3(a,b), respectively, and demonstrate that SI-enhancement can provide depth sectioning of the microspheres to a resolution of 1.2 um, which matches well with our expected axial resolution. No such depth localization is apparent with the conventional WF depth-slice, where QP signal from the microspheres propagates through all depth slices. We note in Fig. 3c that, although the beads are well localized, a haze of QP signal (indicated by yellow arrows) is present– this is an artifact of the "missing-cone" problem in ODT, arising from incomplete frequency coverage, and is typically dealt with in post-processing[47]. To demonstrate how SI changes the frequency content of the imaged volume, we show in Figs. 3(e,f) the axial profiles of radially-averaged 3D Fourier transforms of the SI-enhanced and WF imaging volumes, respectively. The Fourier transform of the WF imaging volume clearly shows that the imaged spatial frequencies lie on the spherical shell (Ewald cap) associated with coherent imaging. In contrast, the Fourier transform of the SI-enhanced volume depicts the distinct butterfly shape associated with ODT[22,36], which allows optical sectioning and enables 3D QP. As expected, Fig. 3e also shows twice the lateral frequency support as Fig. 3f. Figs. 3(g,h) show zooms of regions in Figs. 3(a,b) outlined in yellow to emphasize the improvement in visualization capability of SI-enhanced QP over WF. Though the microspheres are within the system's diffraction limit and are visible in the conventional WF QP image zoom in regions of sparse microsphere density (indicated with green arrows in Fig. 3h), microsphere edges show significant defocus as well as susceptibility to diffraction artifacts. In regions of high microsphere density (indicated with yellow arrows in Fig. 3h), these defocus and diffraction artifacts can effectively hinder clear visualization of individual microspheres. In the SI QP image zoom, such defocus and diffraction artifacts are effectively sectioned out and result in clear and sharp visualization of all individual microspheres. Supplementary Fig. 7 rigorously demonstrates that, even with sample features well within the diffraction limit, WF QP imaging is susceptible to diffraction artifacts from defocus – conversely, SI QP imaging clearly demonstrates its tomographic capability to strongly section out defocused artifacts.

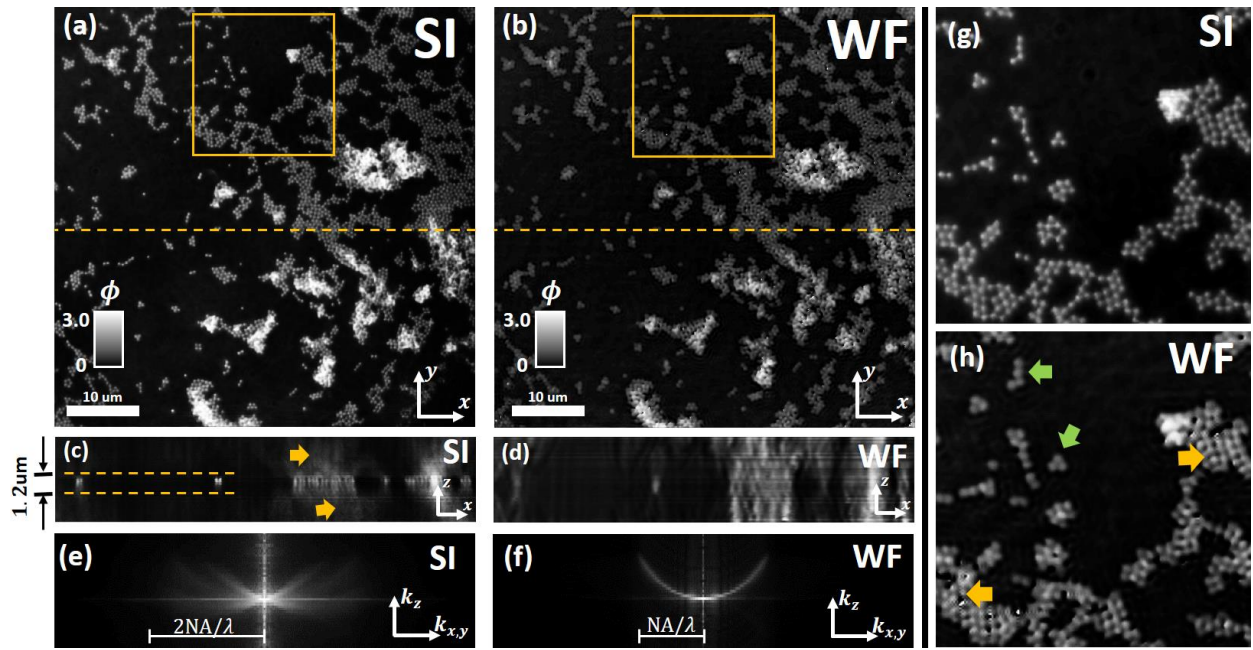

**Figure 3 | Demonstrating SI's capabilities for enabling 3D quantitative-phase (QP) imaging. (a,b)** We compare SI-enhanced and conventional widefield (WF) QP imaging volumes when visualizing 520nm diameter microspheres. Clear improvements allowed by SI include increased lateral resolution and depth-localization. **(c,d)** Axial cross-sections from the SI-enhanced and WF imaging volumes (location marked by dashed yellow line in (a,b)) show the microspheres depth-localized to a resolution of 1.2um with SI-enhancement, while no depth-localization is apparent with conventional WF. **(e,f)** The axial cross-sections of radially-averaged Fourier transforms show the 3D frequency content of the SI-enhanced and WF imaging volumes. The Fourier transform of the WF imaging volume clearly shows the Ewald cap associated with conventional coherent imaging. In contrast, the Fourier transform of the SI-enhanced imaging volume depicts the distinct butterfly shape associated with ODT. **(g,h)** Zooms of the region outlined in yellow from (a,b), respectively, are shown. The 520nm diameter beads fall just within the bound set by the diffraction limit, and so are theoretically resolvable – however, coherent noise and out-of-focus diffraction artifacts deem sections of the zoom (indicated by yellow arrows in (h)) practically irresolvable without the enhancements allowed by SI.

## B. Multimodal 3D sub-diffraction resolution biological visualization.

SI 3D subdiffraction cellular resolution has been demonstrated for fluorescent imaging[30,48] – here, however, we experimentally demonstrate for the first time to our knowledge, SI being used for 3D subdiffraction imaging of both QP and fluorescence in a single, multimodal, optical system. The technical design of this system is detailed in Methods and Supplementary Fig. 4, and mainly consists of a SI-DPM add-on module to a conventional SIM system. For this system, we used an imaging objective with NA of 1.3 to image through immersion media with refractive index $n = 1.51$. Our excitation light remained the original broadband, single-mode, illumination at $\lambda_{exc} = 488 \pm 15$ nm. This yields QP lateral diffraction limits of $d_{\parallel,QP} = \lambda_{exc}/\text{NA} \approx 375$nm and a QP tomographic axial diffraction limit of $d_{\perp,QP} = \lambda_{exc}/[n(1 - \cos\theta)] \approx 635$nm. Our fluorescent filter is designed to pass fluorescent emission at $\lambda_f = 545 \pm 20$ nm. Thus, our expected fluorescent lateral and axial diffraction limits are $d_{\parallel,F} = \lambda_f/2\text{NA} \approx 210$nm and $d_{\parallel,F} = \lambda_f/[n(1 - \cos\theta)] \approx 735$nm, respectively.

To demonstrate imaging performance in a biological sample, we fluorescently labelled A549 cells with AlexFluor-488 phalloidin for F-actin visualization, and imaged QP and fluorescence (Fig. 4 and Supplementary Figs. 8 and 9). While QP imaging offers endogenous mass-based contrast, it lacks sensitivity to specific organelles and cellular components. Therefore, coupling QP imaging with fluorescence allows for the simultaneous evaluation of mass and other descriptors of specific cytological components, and could be further used to delineate organelle boundaries for the determination of refractive index. For both modalities, SI imaging offered dramatic visualization enhancements when compared to conventional WF counterparts. We begin

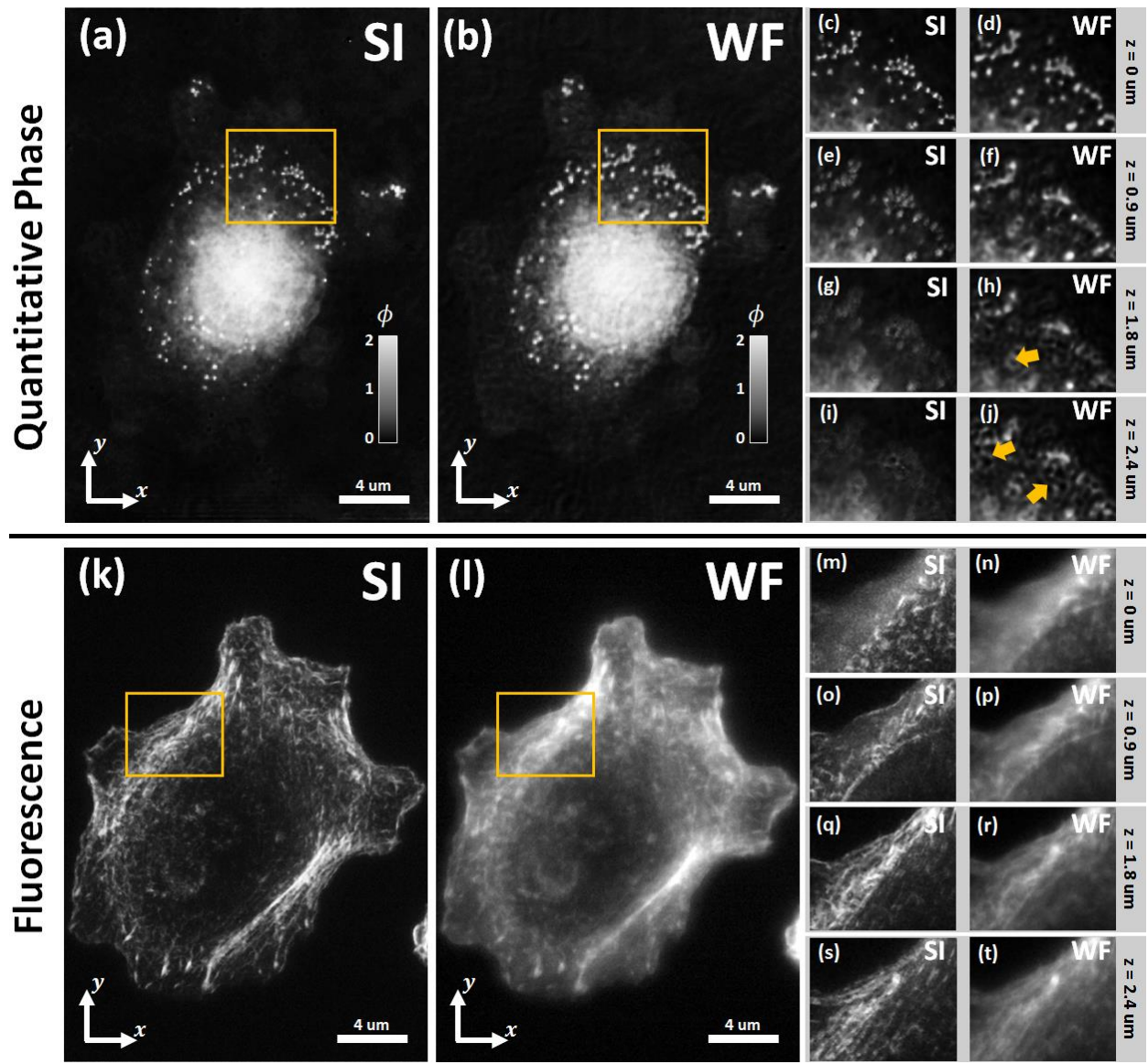

**Figure 4 | 3D visualization capabilities of QP and fluorescent imaging. (a,k)** SI and **(b,l)** WF imaging performances are compared when visualizing an individual A549 cell. SI-enhanced QP sub-diffraction resolution is clearly evident from a zoom of the region outlined in yellow in (a,b), which shows several high phase-delay structures clearly resolvable with (c) SI but not (d) WF QP imaging. 3D imaging capabilities between SI and WF are compared when considering defocused sample planes through the **(c,e,g,i)** SI and **(d,f,h,j)** WF image volumes. In the SI volume, the sharp QP signal from the high phase-delay structures attenuated with increasing defocus, indicating optical depth sectioning. In contrast, the WF volume showed the QP signal from the high phase-delay structures diffracting out into the defocused planes, leading to diffraction artifacts indistinguishable from in-focus QP signal. Fluorescent resolution was also enhanced when comparing **(k)** SI to **(l)** WF imaging, respectively. **(m,o,q,s)** Defocused planes show that SI fluorescence imaging demonstrates clear optical sectioning and shows the actin morphology undergoing clear organizational changes through different depths of the cell. In contrast, defocused planes through the **(n,p,r,t)** WF fluorescence volume show a strong defocused signal throughout the volume stack, which hinders visualization of important high-resolution features.

by comparing imaging performance between SI and WF QPM for an individual A549 cell (Fig. 4(a,b)). We zoom into a region above the nucleus to visualize a cluster of mass localizations. Molecular labelling would be required to truly ascertain the identity of these high phase-delay structures. However, given their small size and high phase-delay, we hypothesize that these are small lipid based vesicles which are known to have a high refractive index lipid bilayer, leading to the relatively high phase delay for the small object[49]. A previous work utilized total internal reflectance microscopy (TIRF) to monitor the secretion of ATP-containing vesicles from the same cell line, and reported these vesicles to surround the perinuclear region[50]. As seen in Figs 4(a,b) the visible high phase-delay vesicles indeed surround the apical portion of the cell where the nucleus likely resides. From the SI and WF zooms in Fig 4(c,d), respectively, we see that these QP structures are beyond the diffraction limit – the factor of 2 resolution enhancement enabled by SI, however, allows clear visualization of the individual localizations (Supplementary Fig. 9(d) quantitatively demonstrates adjacent localizations to have QP peak-to-peak distances of ~230 nm). Furthermore, Figs. 4(c-j) demonstrate that axial translation of the sample results in severe diffraction artifacts from the out-of-focus localizations in the conventional WF visualization (indicated by yellow arrows in Figs. 4(h,j)). Depending on the specific features in the sample, these artifacts may even be indistinguishable from in-focus QP signal. With SI enhancement, however, the localizations effectively disappear from out-of-focus QP images. This phenomenon attributes to the 3D sectioning ability that SI enables in QP imaging by filling out axial frequency space

The F-actin visualization was also drastically improved when comparing fluorescent SI super-resolution to diffraction-limited WF imaging (Figs. 4(k,l)). As can be seen, SI was required to visualize individual actin units. Previous work has demonstrated that 3-beam fluorescent SI results in 3D resolution gain as well as out-of-focus rejection via filling of the missing cone[30]. We experimentally confirm this by showing a zoom region (outlined in yellow in Figs. 4(k,l)) undergoing defocus (Figs. 4(m-t)). Increased axial visualization is apparent when comparing the defocused zooms of SI to conventional WF. Defocused signal is abundant in the WF zooms and occludes clear visualization of the important high-frequency content associated with the F-actin – conversely, the SI-enhanced zooms demonstrate clear optical sectioning, and the imaged actin morphology shows distinct changes as the cells were axially scanned through the image focus. Supplementary Fig. 9(h) quantitatively indicates improvement in lateral resolution by an intensity profile drawn across an in-focus actin filament (Supplementary Fig. 9(g)) – the width of the filament was 180 nm between the signal troughs in the SI fluorescence image, while no resolvable width could be measured in the WF image.

# V. Discussion.

From a theoretical perspective, this work (1) introduced the framework for SI 3D sub-diffraction QP imaging, (2) drew parallels between SI-QPM and oblique illumination microscopy, and (3) equated SI-QPM to a multiplexed form of the more established ODT. However, more fundamentally, this work conveys that SI constitutes a sub-diffraction resolution technique compatible with both fluorescence and QP.

This multimodal compatibility has direct implications that can benefit microscopy for biological research. QP imaging provides a technology with which to noninvasively analyze endogenous cellular biophysical and biochemical parameters. It has shown promise in studying whole-cell spectroscopy, morphology, mass, stiffness, and refractive-index/optical-path-length distributions [19-23]. Furthermore, QP is effectively free and fast – no major sample preparation procedures are necessary to get strong imaging signal and camera integration times rarely pose a problem for longitudinal biological studies. However, QP imaging inherently has no molecular specificity, and the biological parameters extracted from QP cannot be localized to specific cellular components. This prevents analysis of localized and structure-specific QP, which can hinder studies exploring morphology, mechanics, mass, spectra, or density endogenous to individual sub-cellular components. Towards this end, due to its ability for molecular-specific contrast, fluorescence microscopy directly complements QP imaging. Fluorescence microscopy is the dominant imaging choice when studying interactions and dynamics of specific sub-cellular components, as is necessary in studies of gene expression, protein localization, intracellular transport, organelle dynamics, diffusion kinetics, etc[13-18]. A single optical system that incorporates QP and fluorescent imaging can allow registered, multimodal visualization of specific, localized regions of the sample and perhaps enable molecular-specific quantitative analysis. Indeed, a few past works may have considered this potential and introduced optical systems that efficiently combined both modalities[51,52]. Unfortunately, though such systems may represent important tools for biological analysis, they remain diffraction-limited, and a generalized scheme to achieve robust sub-diffraction resolution imaging with coherent and fluorescent contrast has remained elusive.

As shown in this work, SI represents a solution to this problem. Both coherent and fluorescent imaging have individually showed that a factor 2 resolution gain dramatically enhances biological visualization. Fig. 4 and Supplementary Figs. 8 and 9 demonstrate this with QP imaging by significantly improving visualization of high phase-delay structures surrounding the nuclear region of the cell. ODT also exploits this factor 2 resolution gain when visualizing 3D refractive-index maps at sub-diffraction resolutions[22]. In fluorescence imaging, previous applications utilizing SI-enabled resolution-doubling have included super-resolved visualization of microtubule dynamics associated with the polymerization of α-tubulin in Drosophila melanogaster S2 cells[48]. Mitochondria dynamics in HeLa cells were also imaged, and SI-enabled super-resolution was required to visualize the fine mitochondrial features associated with the cristae, usually seen only with electron microscopes.

We envision that biological insights extracted from such resolution enhancements from typically separate coherent and fluorescent imaging systems, can be potentially combined within the SI framework. Such a synergy may be important to comprehensively study biological components with distinct molecular and biophysical/biochemical functions. For example, cell cytoskeleton, known to be a network of F-actin, microtubules, and other intermediate filaments connecting most cellular structures, directly affects the mechanical properties of the cell. Previous studies have explored the distributions of mechanical stresses and displacements within the cell in response to applied loads and have modelled the cytoskeleton as a network of discrete, stressed elements[53,54]. The cytoskeleton, however, also affects various important signal transduction pathways (STP) that control how information (from either mechanical or molecular stimuli) is passed from the cell membrane to structures within the cytoplasm or nucleus. Important cytoskeletal components that enable this include molecules from the glycolytic enzyme, protein kinase, lipid kinase, hydrolase, and GTPase families[55]. Multimodal QP/fluorescent sub-diffraction imaging can offer a unique ability to probe molecular and mechanical interactions in the cytoskeleton during STP events by enabling visualization of molecular-specific mechanical forces as well as whole-cell mechanical responses to signal transduction events. Applications of this can be important to explore how mechanical or mechanotransduction events affect gene expression and protein synthesis, as well as whole-cell functions such as cell growth, differentiation, locomotion, cytokinesis, and apoptosis.

**Acknowledgements**
We acknowledge the generous financial support provided by the National Science Foundation (CBET-1403905)

**Author Contributions**
S.C. designed the microscope, prepared calibration samples, performed imaging experiments, and analyzed/reconstructed image data. W.J.E. prepared cellular samples for biological imaging. A.W. and J.A.I. directed and supervised the project.

**Competing financial interests:** The authors declare no competing financial interests

# Supplementary Information

## Supplementary Note 1

In this section, we rigorously adapt the existing mathematical framework for 3D super-resolution fluorescent imaging via SI to 3D sub-diffraction resolution QP imaging. Though we highlight distinct differences between the two frameworks, we strive to maintain similar notation and derivation strategies as those presented in the original work by Gustafsson *et al*[30]. We hope that such similarities may encourage readers to draw comparisons between the mathematical frameworks for SI-enabled sub-diffraction QP and super-resolution fluorescent 3D reconstructions. We start by briefly reviewing how spatial frequencies propagate through a coherent imaging system.

**Coherent propagation of spatial frequencies through system aperture**

For widefield coherent imaging, the illumination beam is typically considered to be a monochromatic plane-wave of wave-vector $\boldsymbol{k}_{\text{illum}} = (0,0, k_\lambda)$ incident with a flat wavefront on a diffracting sample (shown below in Supplementary Figure 1(a)). Here, $k_\lambda = 2\pi/\lambda$ is the wave-vector magnitude, and $\lambda$ is the illumination wavelength. The interaction of the sample with the illumination beam results in a total diffraction that can be decomposed into plane-wave components, each with a differently oriented wave-vector. Because $\lambda$ is maintained across all plane-wave components, these differently oriented wave-vectors share the common wave-vector magnitude of $k_\lambda$. Thus, in 3D Fourier space, the set of all diffracted wave-vectors trace a sphere (i.e., Ewald sphere) of radius $k_\lambda$ centered at the origin.

We denote $\boldsymbol{k}_s$ as a wave-vector from the set of plane-wave components accepted through the numerical aperture of the imaging objective. The set of all possible wave-vectors $\boldsymbol{k}_s$ traces a spherical cap on the Ewald sphere (illustrated below by a circular arc in Supplementary Figure 1(b)), and describes the 3D spatial frequencies of the component plane-waves that are allowed to propagate to the imaging detector. In the lightly-scattering approximation, valid for largely transparent samples, $\boldsymbol{k}_s \approx \boldsymbol{k}_{\text{obj}} + \boldsymbol{k}_{\text{illum}}$, where $\boldsymbol{k}_{\text{obj}}$ is a 3D spatial frequency inherent in the sample. Thus, we see that $\boldsymbol{k}_{\text{obj}}$ is simply shifted from $\boldsymbol{k}_s$ by the constant illumination wave-vector, $\boldsymbol{k}_{\text{obj}} \approx \boldsymbol{k}_s - \boldsymbol{k}_{\text{illum}}$, for the set of all detected $\boldsymbol{k}_s$. In Fourier space, the set of all $\boldsymbol{k}_{\text{obj}}$ thus traces a spherical cap coincident with the origin (shown below in Supplementary Figure 1(c) in the case of widefield imaging), and denotes the region of the sample's 3D spatial-frequency spectrum that can be imaged. Thus, this region is considered the coherent system's transfer function (TF). The aim of sub-diffraction resolution imaging is to capture more sample spatial frequencies than those directly encompassed by this TF.

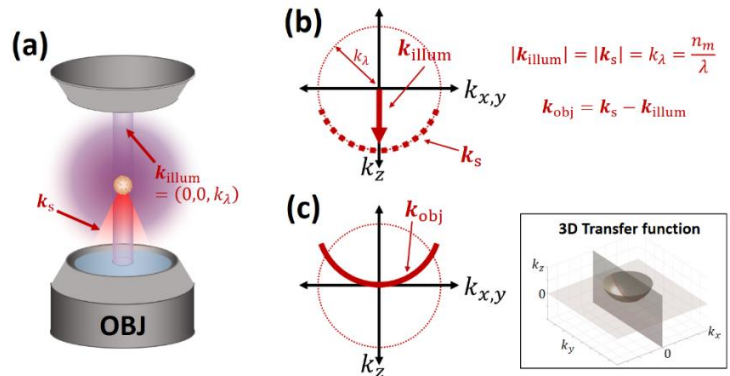

**Supplementary Figure 1.** Illustration depicting transfer of spatial frequencies in a coherent imaging system. (a) Illustration depicting imaging setup for orthogonal, widefield, coherent illumination. (b) Fourier diagram illustrating the illumination wave-vector, the Ewald sphere (in dashed outline) for all possible diffracted spatial frequencies, as well as the set of all diffracted spatial frequencies $\boldsymbol{k}_s$ that can propagate to the imaging detector. (c) Fourier diagram illustrating the coherent system's transfer function as the region of the sample's 3D spatial-frequency spectrum that give rise to the set of all detected spatial frequencies $\boldsymbol{k}_s$.

### 3D image formation with axial scanning of sample

The imaging formation process in a coherent imaging system with complex electric-field imaging can be described with standard properties of linearity and translation-invariance. Namely, the electric-field of the system's image is a blurred version of the electric-field diffracting through the sample. Mathematically, this can be described by:

$$y(\boldsymbol{r}) = h(\boldsymbol{r}) \otimes [s(\boldsymbol{r}) \cdot i(\boldsymbol{r})] \qquad (1)$$

where $\boldsymbol{r} = (x, y, z)$ is the 3D spatial coordinate vector, $s(\boldsymbol{r})$ is the sample's complex transmittance function, $i(\boldsymbol{r})$ is the illumination electric-field through the sample, $y(\boldsymbol{r})$ is the acquired image's complex electric-field (obtained in this manuscript via off-axis holography), $h(\boldsymbol{r})$ is the imaging system's coherent point-spread-function (PSF), and $\otimes$ is the convolution operator.

We note here that $h(\boldsymbol{r})$ describes the system's PSF for the spatial frequencies present in the *diffraction propagating* from the sample that are able to arrive at the imaging detector, not simply the inherent spatial frequencies in the sample. Thus, for the purposes of this derivation, we label the Fourier transform of $h(\boldsymbol{r})$ as the system's propagating transfer function (pTF), such that $h(\boldsymbol{r}) \overset{\text{FFT}}{\iff} H_p(\boldsymbol{k})$, where $\boldsymbol{k} = (k_x, k_y, k_z)$ is the 3D spatial frequency vector and $H_p(\boldsymbol{k})$ is the system pTF that encompasses the region of 3D Fourier space that describe the spatial frequencies present in the diffracted waves from the sample. Mathematically, the pTF is exactly the set of all possible detected wave-vectors $\boldsymbol{k}_s$ (shown above in Supplementary Figure 1(b)).

In the specific case where the illumination 3D electric-field can generally be written as a sum of discrete components that can be separated into axial and lateral harmonic functions:

$$i(\boldsymbol{r}_{x,y}, z) = \sum_m c_m \, i_m(z) \cdot j_m(\boldsymbol{r}_{x,y}) \qquad (2)$$

where $\boldsymbol{r}_{x,y} = (x, y)$ is the 2D lateral spatial coordinate vector, $i_m(z)$ is the $m^{\text{th}}$ axial complex harmonic function, and $j_m(\boldsymbol{r}_{x,y})$ is the $m^{\text{th}}$ lateral complex harmonic function, and $c_m$ is an arbitrary coefficient. If this illumination function is fixed with respect to the sample during image acquisition, the expression for the 3D volumetric image can be simply written by substituting Eq. (2) into Eq. (1) and expanding the convolution integral:

$$\begin{aligned} y(\boldsymbol{r}) &= [h \otimes (s \cdot i)](\boldsymbol{r}) \\ &= h(\boldsymbol{r}) \otimes \left[ s(\boldsymbol{r}) \sum_m c_m \, i_m(z) \cdot j_m(\boldsymbol{r}_{x,y}) \right] \\ &= \sum_m c_m \cdot \int h(\boldsymbol{r} - \boldsymbol{r}') \, s(\boldsymbol{r}') \, i_m(z') \, j_m(\boldsymbol{r}'_{x,y}) \, d\boldsymbol{r}' \end{aligned} \qquad (3)$$

where $\boldsymbol{r}' = (x', y', z')$ refers to the 3D spatial coordinate vector in the reference frame of the sample. Importantly, in the case of a fixed illumination function with respect to the sample during image acquisition, Eq. (3) shows the axial harmonic function depending on *only* the sample's reference coordinates. Conversely, if the sample was axially *scanned* during image acquisition, the axial harmonic function would instead depend on the *difference* between the sample's and image's coordinate frames, as first importantly noted by Gustafsson *et al*[30]. Thus, in the case of axial scanning of the sample, the integral form of Eq. (3) is slightly modified:

$$y(\mathbf{r}) = \sum_m c_m \cdot \int h(\mathbf{r} - \mathbf{r}') \, i_m(z - z') \, s(\mathbf{r}') \, j_m(\mathbf{r}'_{x,y}) \, d\mathbf{r}'$$

$$= \sum_m c_m \cdot \left[ \left( h(\mathbf{r}) \cdot i_m(z) \right) \otimes \left( s(\mathbf{r}) \cdot j_m(\mathbf{r}_{x,y}) \right) \right] \quad (4)$$

$$= \sum_m c_m \cdot y_m(\mathbf{r})$$

where $y_m(\mathbf{r})$ designates the $m^{\text{th}}$ term of the $y(\mathbf{r})$. Fourier transforming $y_m(\mathbf{r})$, we get:

$$Y_m(\mathbf{k}) = H_m(\mathbf{k}) \cdot \left[ S(\mathbf{k}) \otimes J_m(\mathbf{k}_{x,y}) \right] \quad (5)$$

where $S(\mathbf{k}), J_m(\mathbf{k}_{x,y})$, and $I_m(k_z)$ are the Fourier transforms of $s(\mathbf{r}), j_m(\mathbf{r}_{x,y})$, and $i_m(z)$, respectively, $\mathbf{k}_{x,y} = (k_x, k_y)$ is the 2D lateral spatial frequency vector, and $H_m(\mathbf{k}) = [H_p \otimes I_m](\mathbf{k})$ is the Fourier transform of $[h \cdot i_m](\mathbf{r})$.

Because the above mathematical treatment fundamentally relies only on modelling the imaging process as a linear and translation-invariant process (a condition satisfied by coherently imaging complex electric-fields), the final imaging expression Eq. (5) is identical to that presented in Eq. (6) in Gustafsson *et al*[30], where linearity and translation-invariance requirements were fulfilled by imaging intensity through a fluorescent (incoherent) system. We now apply Eq. (5) to specifically describe coherent 3D imaging under widefield and SI conditions.

### 3D image formation with orthogonal widefield illumination and axial scanning of sample

In the case of coherent widefield illumination, which consists of illuminating the sample with a flat wavefront, the illumination beam can be described simply as an axially propagating plane-wave, $i_{WF}(\mathbf{r}) = \exp(j\, k_\lambda z)$. Comparing this to the expression in Eq. (2), we clearly see that $i_{WF}(\mathbf{r})$ has only the $m = 0$ component and can be separated into axial and lateral harmonic functions $i_{WF}(\mathbf{r}) = i_0(z) \cdot j_0(\mathbf{r}_{x,y})$ where $i_0(z) = \exp(j\, k_\lambda z)$ and $j_0(\mathbf{r}_{x,y}) = 1$, respectively, and coefficient $c_0 = 1$. Fourier transforming $i_0(z)$ and $j_0(\mathbf{r}_{x,y})$ results in $I_0(k_z) = \delta(k_z - k_\lambda)$ and $J_0(\mathbf{k}_{x,y}) = \delta(\mathbf{k}_{x,y})$. After substitution back into Eq. (5), we see that the 3D image formed under widefield illumination with axial scanning of the sample is $Y_0(\mathbf{k}) = H_0(\mathbf{k}) \cdot S(\mathbf{k})$, where $H_0(\mathbf{k}) = [H_p \otimes I_0](\mathbf{k}) = H_p(\mathbf{k} - \mathbf{k}_\lambda)$ and $\mathbf{k}_\lambda = (0,0,k_\lambda)$.

We note now that $H_p(\mathbf{k} - \mathbf{k}_\lambda)$ encompasses exactly the same region of Fourier space as the system TF, illustrated above in Supplementary Figure 1(c). Thus, $Y_0(\mathbf{k})$ is shown to be exactly a low-pass filtered version of $S(\mathbf{k})$ with the system TF. Though this result is expected, it demonstrates the validity of Eq. (5) in describing the coherent image formation process. We now use Eq. (5) to mathematically describe the image formation process under structured illumination.

### 3D image formation with 3-beam sinusoidal illumination and axial scanning of sample

In this manuscript, we achieve structured illumination of the sample with three mutually coherent tilted plane-waves. Thus, the 3D illumination electric-field can be mathematically expressed as:

$$i_{SI}(\mathbf{r}) = \exp(j\, (\mathbf{k}_0 \cdot \mathbf{r} + \phi_0)) + \exp(j\, (\mathbf{k}_1 \cdot \mathbf{r} + \phi_1)) + \exp(j\, (\mathbf{k}_2 \cdot \mathbf{r} + \phi_2)) \quad (6)$$

where $\mathbf{k}_0, \mathbf{k}_1, \mathbf{k}_2$ and $\phi_0, \phi_1, \phi_2$ are the wave-vectors and reference phases for the three component plane waves, respectively.

In the specific case where 3-beam interference is achieved by interfering only the $0^{th}$ and $\pm 1^{st}$ orders diffracting from a sinusoidal patterned structured element with 2D spatial frequency vector $\mathbf{k}_T$, we can impose the additional constraints $\mathbf{k}_0 = (0,0,k_\lambda)$, $\mathbf{k}_1 = (\mathbf{k}_T, k_{T,z})$, $\mathbf{k}_2 = (-\mathbf{k}_T, k_{T,z})$, $\phi_0 = 0$, and $\phi_1 = -\phi_2 = \phi$, where $\mathbf{k}_T$ and $k_{T,z}$ are the projections of $\mathbf{k}_1$ onto the 2D lateral $\mathbf{k}_{x,y}$ and 1D axial $k_z$ coordinate spaces, respectively. Because $\lambda$ is maintained throughout the imaging process, $\mathbf{k}_0, \mathbf{k}_1, \mathbf{k}_2$ share the same wave-vector magnitude, ie. $|\mathbf{k}_0|^2 = |\mathbf{k}_1|^2 = |\mathbf{k}_2|^2 = |\mathbf{k}_T|^2 + |k_{T,z}|^2 = k_\lambda^2$. Thus, we can rewrite Eq. (6) as:

$$\begin{aligned} i_{SI}(\mathbf{r}) &= \exp(j\, k_\lambda z) + \exp\left(j\left(\mathbf{k}_T \cdot \mathbf{r}_{x,y} + k_{T,z} + \phi\right)\right) \\ &\quad + \exp\left(j\left(-\mathbf{k}_T \cdot \mathbf{r}_{x,y} + k_{T,z} - \phi\right)\right) \\ &= \exp(j\, k_\lambda z) + e^{j\phi} \exp(j\, \mathbf{k}_T \cdot \mathbf{r}_{x,y}) \exp(j\, k_{T,z} \cdot k_z) \\ &\quad + e^{-j\phi} \exp(-j\, \mathbf{k}_T \cdot \mathbf{r}_{x,y}) \exp(j\, k_{T,z} \cdot k_z) \end{aligned} \tag{7}$$

Comparing the final form of Eq. (7) with Eq. (2), we see that $i_{SI}(\mathbf{r})$ is composed of the $m = 0,1,2$ components with coefficients $c_0 = 1$, $c_1 = e^{j\phi}$, and $c_2 = e^{-j\phi}$, respectively. Lateral and axial harmonic functions with associated Fourier expressions are shown below:

$$\begin{aligned} i_0(z) = \exp(j\, k_\lambda z) &\overset{\text{FFT}}{\Longleftrightarrow} I_0(k_z) = \delta(k_z - k_\lambda) \\ j_0(\mathbf{r}_{x,y}) = 1 &\overset{\text{FFT}}{\Longleftrightarrow} J_0(\mathbf{k}_{x,y}) = \delta(\mathbf{k}_{x,y}) \\ i_1(z) = \exp(j\, k_{T,z} \cdot k_z) &\overset{\text{FFT}}{\Longleftrightarrow} I_1(k_z) = \delta(k_z - k_{T,z}) \\ j_1(\mathbf{r}_{x,y}) = \exp(j\, \mathbf{k}_T \cdot \mathbf{r}_{x,y}) &\overset{\text{FFT}}{\Longleftrightarrow} J_1(\mathbf{k}_{x,y}) = \delta(\mathbf{k}_{x,y} - \mathbf{k}_T) \\ i_2(z) = \exp(j\, k_{T,z} \cdot k_z) &\overset{\text{FFT}}{\Longleftrightarrow} I_2(k_z) = \delta(k_z - k_{T,z}) \\ j_2(\mathbf{r}_{x,y}) = \exp(-j\, \mathbf{k}_T \cdot \mathbf{r}_{x,y}) &\overset{\text{FFT}}{\Longleftrightarrow} J_2(\mathbf{k}_{x,y}) = \delta(\mathbf{k}_{x,y} + \mathbf{k}_T) \end{aligned} \tag{8}$$

Substitution of the Fourier expressions in Eq. (8) into Eq. (5) yields the Fourier distributions of the component terms in the acquired 3D image:

$$\begin{aligned} Y_0(\mathbf{k}) &= H_0(\mathbf{k}) \cdot S(\mathbf{k}) \\ H_0(\mathbf{k}) &= H_p(\mathbf{k} - \mathbf{k}_\lambda) \\ \\ Y_1(\mathbf{k}) &= H_1(\mathbf{k}) \cdot S(\mathbf{k} - \mathbf{k}_{T,3D}) \\ H_1(\mathbf{k}) &= H_p(\mathbf{k} - \mathbf{k}_{T,z,3D}) \\ \\ Y_2(\mathbf{k}) &= H_2(\mathbf{k}) \cdot S(\mathbf{k} + \mathbf{k}_{T,3D}) \\ H_2(\mathbf{k}) &= H_p(\mathbf{k} - \mathbf{k}_{T,z,3D}) \end{aligned} \tag{9}$$

Here, we recall from above that $\mathbf{k}_\lambda = (0,0,k_\lambda)$ and we define $\mathbf{k}_{T,3D} = (\mathbf{k}_T, 0)$ and $\mathbf{k}_{T,z,3D} = (0,0,k_{T,z})$ as the analogues of $\mathbf{k}_T$, and $k_{T,z}$ in 3D coordinate space. These terms are introduced purely to maintain notational dimension consistency in the spatial frequency domain arguments expressed in Eq. (9). We also note that $H_1(\mathbf{k}) = H_2(\mathbf{k})$, and are distinguished purely for the sake of association with the proper component terms.

In Supplementary Figure 2(a-c) and (d-f) below, we graphically illustrate the relationship between $I_m(k_z)$, $J_m(k_{x,y})$, and $H_m(\boldsymbol{k})$ and the corresponding $Y_m(\boldsymbol{k})$ terms, respectively, for components $m = 0,1,2$. As is clear in the case of $m = 0$, the axial/lateral Fourier pair $\{I_0(k_z); J_0(k_{x,y})\}$ is exactly equivalent to the Fourier representation of widefield illumination – thus, the sample's spatial frequency information contained in $Y_0(\boldsymbol{k})$ is mathematically equivalent to what is imaged via widefield illumination. However, in the case of $m = 1,2$, we see that the combination of axially translated $H_m(\boldsymbol{k})$ (from convolution of $H_p(\boldsymbol{k})$ with $I_m(k_z)$) and laterally translated $S(\boldsymbol{k})$ (from convolution of $S(\boldsymbol{k})$ with $J_m(k_{x,y})$) results in $Y_m(\boldsymbol{k})$ imaging a region of the sample's spatial frequency spectrum that is axially and laterally offset from the region imaged with widefield imaging. To reconstruct a high-resolution image where these image regions of the sample's spatial frequency spectrum are appropriately synthesized, the component images $Y_m(\boldsymbol{k})$ for $m = 0,1,2$ need to be separated and then appropriately Fourier shifted.

Fourier transforming Eq. (4) shows the Fourier spectrum of a single raw image to be a linear superposition of the Fourier spectra of the individual component images:

$$Y(\boldsymbol{k}) = Y_0(\boldsymbol{k}) + e^{j\phi}Y_1(\boldsymbol{k}) + e^{-j\phi}Y_2(\boldsymbol{k}) \tag{10}$$

As in conventional fluorescent SI, disentangling the component terms in Eq. (10) simply requires acquiring multiple, say $N > 3$, raw images $Y(\boldsymbol{k})$ with different known values of $\phi$ to form a linear system of $N$ equations. Standard linear inversion solvers can then be used to solve for the unknown $Y_0(\boldsymbol{k}), Y_1(\boldsymbol{k})$, and $Y_2(\boldsymbol{k})$ components. Varying $\phi$ is accomplished by simple translation of the sinusoidal patterned structured element. Conventional optical diffraction tomography also acquires exactly these components via sequential oblique illuminations (i.e., illuminating individually with the plane-wave components composing the SI, as expressed in Eq. (6): $\exp(j\,\boldsymbol{k}_0 \cdot \boldsymbol{r}), \exp(j\,\boldsymbol{k}_1 \cdot \boldsymbol{r})$, and $\exp(j\,\boldsymbol{k}_2 \cdot \boldsymbol{r})$, respectively).

Once the component terms are known, $Y_1(\boldsymbol{k})$ and $Y_2(\boldsymbol{k})$ can be digitally Fourier shifted by $-\boldsymbol{k}_{T,3D}$ and $\boldsymbol{k}_{T,3D}$ respectively, such that $Y_1(\boldsymbol{k} + \boldsymbol{k}_{T,3D}) = H_1(\boldsymbol{k} + \boldsymbol{k}_{T,3D}) \cdot S(\boldsymbol{k})$ and $Y_2(\boldsymbol{k} - \boldsymbol{k}_{T,3D}) = H_2(\boldsymbol{k} - \boldsymbol{k}_{T,3D}) \cdot S(\boldsymbol{k})$. As shown below in Supplementary Figure 2(g-i), such shifts reposition the imaged content in $Y_1(\boldsymbol{k})$ and $Y_2(\boldsymbol{k})$ to their true positions in Fourier space. Simple addition of these repositioned terms yields:

$$\begin{aligned} Y_{SI}(\boldsymbol{k}) &= Y_0(\boldsymbol{k}) + Y_1(\boldsymbol{k} + \boldsymbol{k}_{T,3D}) + Y_2(\boldsymbol{k} - \boldsymbol{k}_{T,3D}) \\ &= [H_0(\boldsymbol{k}) + H_1(\boldsymbol{k} + \boldsymbol{k}_{T,3D}) + H_2(\boldsymbol{k} - \boldsymbol{k}_{T,3D})] \cdot S(\boldsymbol{k}) \end{aligned} \tag{11}$$

Eq. (11) demonstrates that the total reconstructed region of the sample's spectrum is encompassed by $H_0(\boldsymbol{k}) + H_1(\boldsymbol{k} + \boldsymbol{k}_{T,3D}) + H_2(\boldsymbol{k} - \boldsymbol{k}_{T,3D})$. As is clear from Supplementary Figure 2(j) below, this region contains more of the sample's spectrum than allowed by typical widefield imaging - however, significant axial portions of the sample's spectrum remain uncovered. To image these regions, the above process is repeated for increments of the 2D spatial frequency magnitude contained in the sinusoidal patterned structured element, $|\boldsymbol{k}_T|$. Recall the deterministic relationship between the magnitudes of the 2D lateral spatial frequency $\boldsymbol{k}_T$ and the 1D axial spatial frequency $k_{T,z}$, i.e., $|\boldsymbol{k}_T|^2 + |k_{T,z}|^2 = k_\lambda^2$. Thus, incrementing $\boldsymbol{k}_T$ will inherently axially increment the sample's region of imaged Fourier spectrum and enable solid sub-diffraction resolution imaging enhancements. The total coverage of 3D Fourier space from incremented $\boldsymbol{k}_T$ is illustrated in the main text Figure 2(c,f).

This procedure, in turn, is repeated for incremented rotations of $\boldsymbol{k}_T$ (i.e., rotating the structured element) to isotropically extend the resolution enhancements. Wiener deconvolution techniques can compensate for uneven weighting due to overlap of individual components in the final summations.

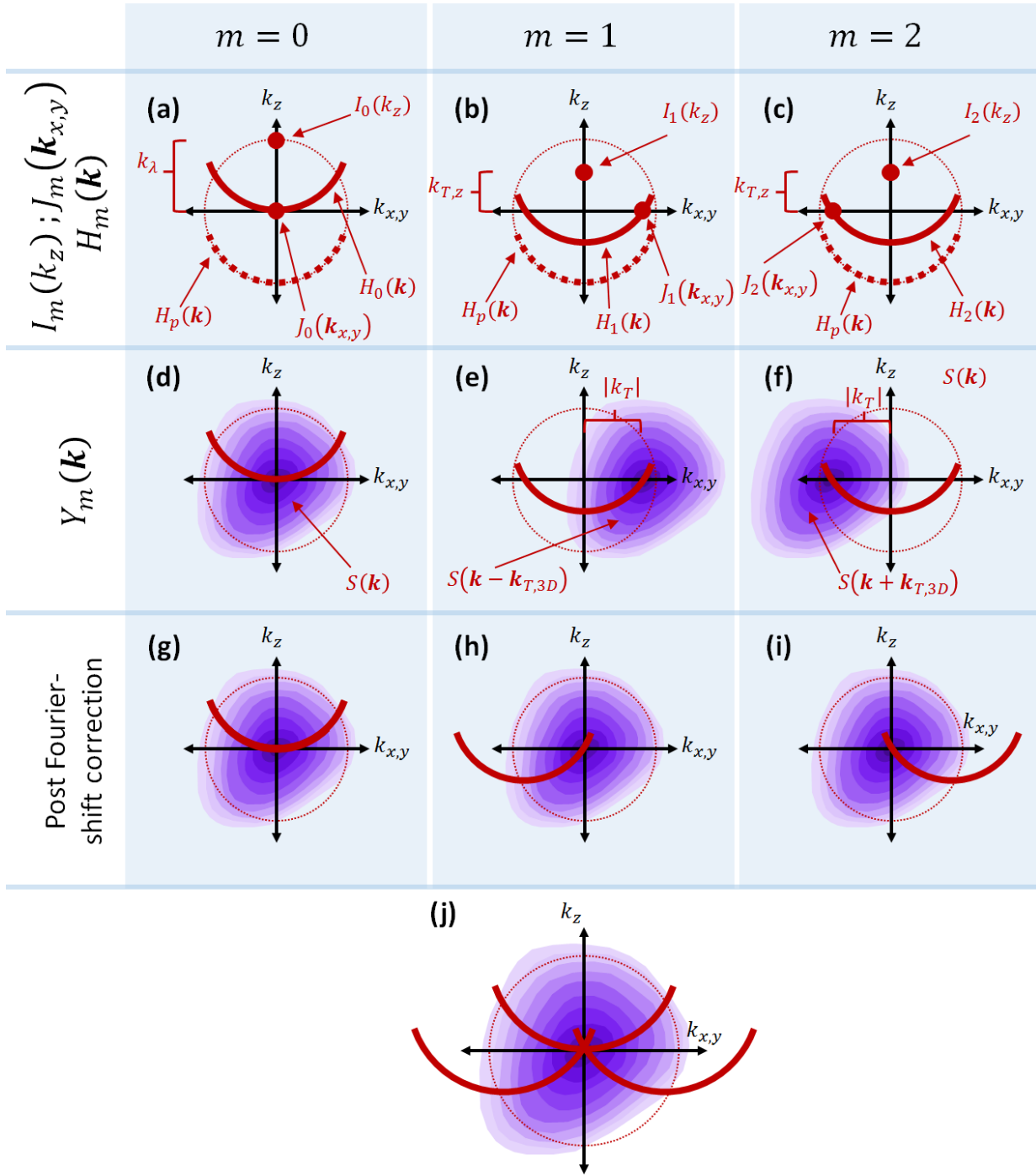

**Supplementary Figure 2.** (a-c) Illustration of the relationship between illumination axial and lateral harmonic components, $I_m(k_z)$ and $J_m(\boldsymbol{k}_{x,y})$ respectively, and $H_m(\boldsymbol{k})$ for component orders $m = 0,1,2$. (d-f) Illustration of the relative positioning between the sample spectrum $S(\boldsymbol{k})$ and the image system's region of accepted spatial frequencies $H_m(\boldsymbol{k})$ for components $Y_0(\boldsymbol{k}), Y_1(\boldsymbol{k}), Y_2(\boldsymbol{k})$. (g-i) Illustrating the regions of accepted sample spectrum after correcting for the original spatial-frequency shifts present in the sample information in components $Y_0(\boldsymbol{k}), Y_1(\boldsymbol{k}), Y_2(\boldsymbol{k})$. (j) Illustrating that the synthesized region of accepted sample spectrum contains more information than allowed by simple widefield coherent imaging (as is illustrated in (d)).

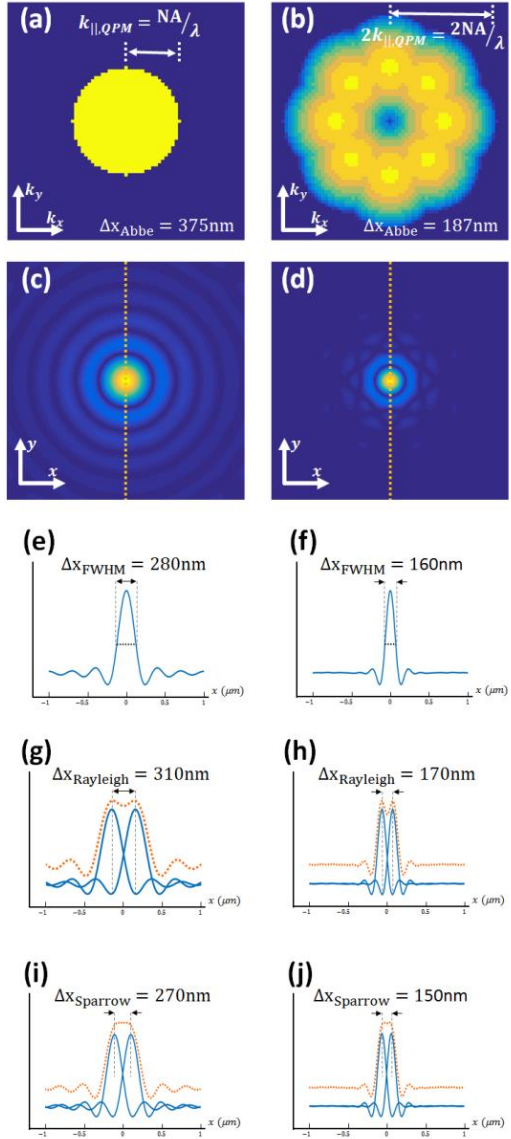
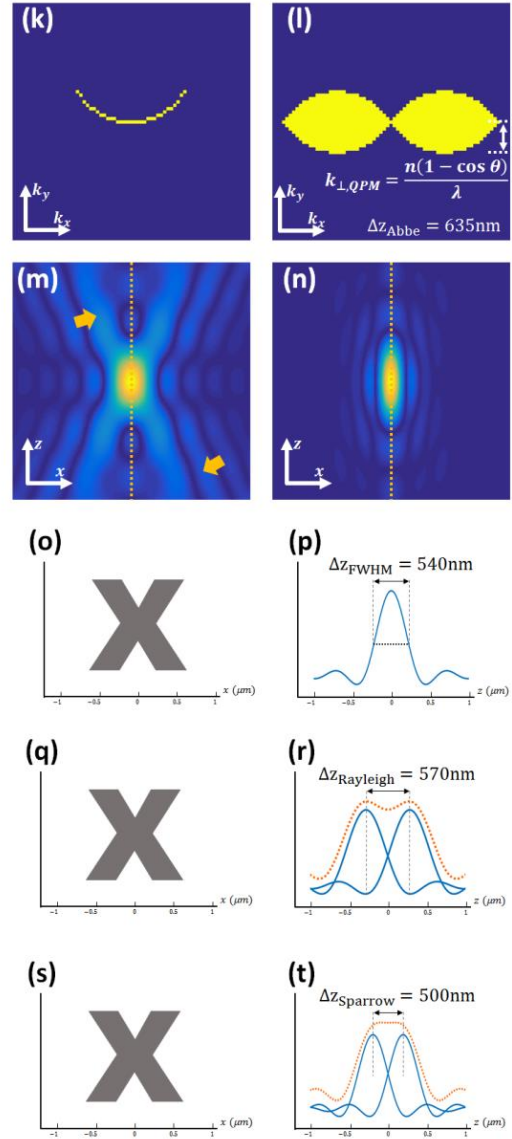

**Supplementary Figure 3.** Conventional resolution metrics are compared for the (a-j) lateral and (k-t) axial point-spread-functions (PSF) for conventional WF and SI QP imaging techniques, simulated under conditions of 1.3 NA imaging through immersion media of refractive index n = 1.51. The sample is assumed to diffract coherently with uniform phase. (a,b) 2D lateral projections of the 3D transfer-functions (TF) for WF and SI QP imaging, respectively, show the lateral Abbe resolution limits, which equates the minimum resolvable spatial-harmonic period to the reciprocal of the frequency bounds of the TFs, to be $\lambda/NA$ = 375nm and $\lambda/2NA$ = 187nm, respectively. (l) Axial cuts of the SI QP TF (assuming axial space has been completely filled out) show the corresponding axial Abbe resolution limit is 635nm. 3D complex PSFs were calculated via 3D Fourier Transform from the 3D TFs and *magnitude* (c,d) lateral and (m,n) axial cross-sections are shown for WF and SI QP imaging. Note that the (k) WF QP TF shows infinitesimal axial frequency support – hence the concept of WF QP's axial "resolution" is inapplicable. This is demonstrated by the significant non-localized signal (indicated by yellow arrows) in its magnitude PSF's axial cross-section. We show the (e,f,p) *amplitude* line profiles from the regions in dashed yellow in (c,d,n), and compare the resolutions as would be reported by three relevant resolution metrics – (e,f,p) full-width-half-max, (g,h,r) Rayleigh, and (i,j,t) Sparrow. For the coherent Rayleigh and Sparrow metrics, the two point-sources that are required to be "resolvable" are assumed to be diffracting constructively and with equal intensity. As is evident, the reported resolution is highly dependent on the choice of resolution metric (can lead to >100nm difference in reported resolutions), and should not be confused with an actual difference in the system's PSF or imaging performance.

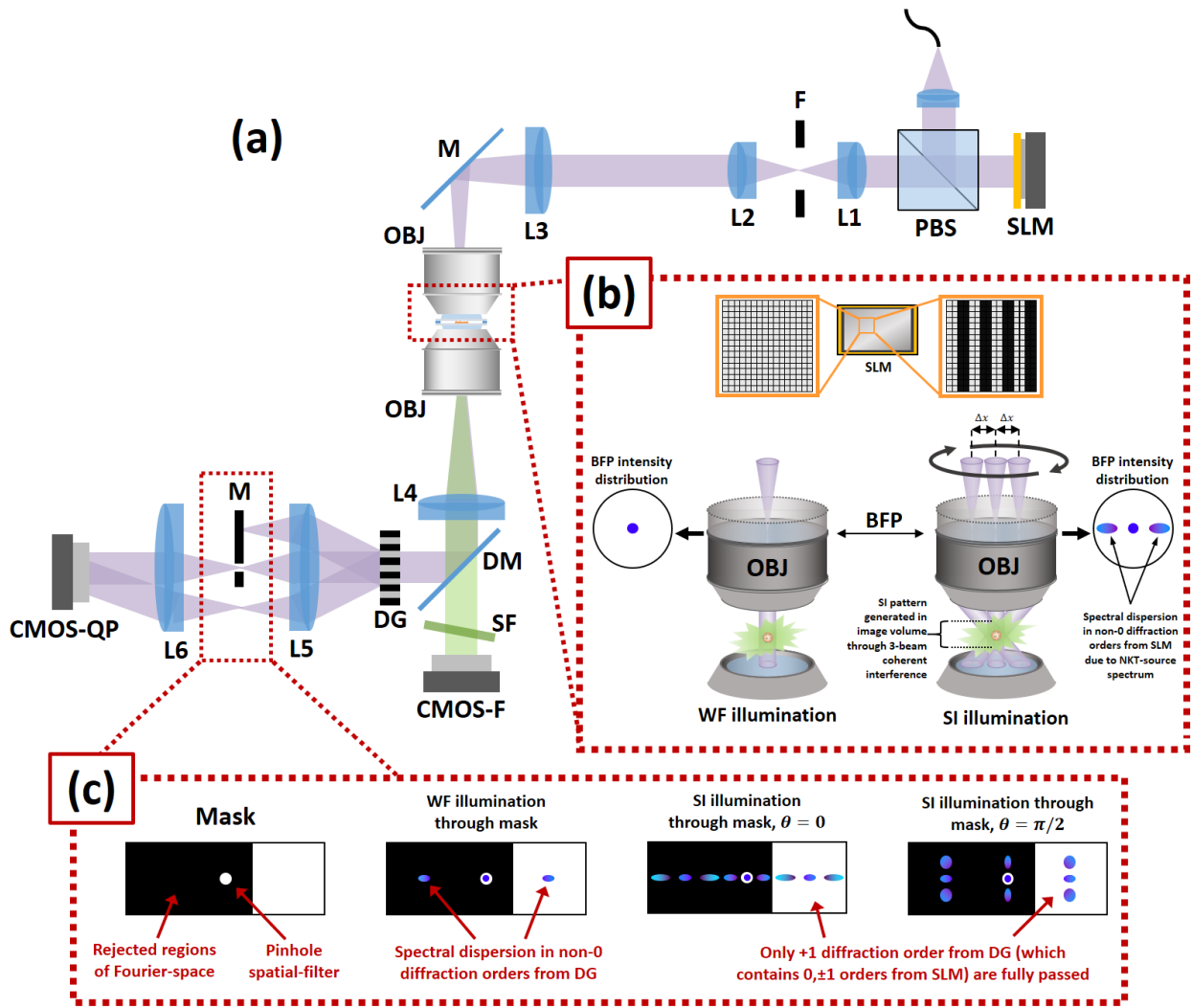

**Supplementary Figure 4.** (a) Optical system diagram is shown, and consolidates conventional fluorescent structured illumination (SI) microscopy with SI diffraction phase microscopy (SI-DPM) for sub-diffraction resolution quantitative-phase (QP) and fluorescent imaging. (b) Conventional widefield illumination is achieved when all the SLM pixels are turned 'ON', resulting in a single focused spot at the back focal plane (BFP) of the imaging objective, and thus a single collimated beam illuminating the sample. To achieve sinusoidal structured illumination, a the SLM is programmed to display a sinusoidal pattern. An adjustable iris diaphragm (F) is placed in the Fourier plane of the first 4f system after the SLM to physically block extraneous diffraction orders resulting from the SLM's pixilation, and ensure the passage of only the $0^{th}$ and $\pm 1^{st}$ diffraction orders from the programmed sinusoidal pattern. This results in three focused spots at the BFP of the imaging objective and a three beam coherent interference through the imaging volume. Due to the broadband spectrum of the NKT-illumination source, the focused spots at the imaging objective's BFP corresponding to the ±1 diffraction orders from the SLM are spectrally dispersed. The distance between the orders (labelled as $\Delta x$) at the BFP directly relates to the spatial frequency of the SI pattern. Rotating the sinusoidal pattern written to the SLM rotates the ±1 diffraction orders around the $0^{th}$ diffraction order and results in rotating the SI pattern through the imaging volume to fill out the sample's frequency space. (c) The diffracted excitation beam from the sample is spectrally separated from the fluorescence, and is fed through a diffraction-phase-microscope (DPM). A Ronchi grating (DG) is positioned at the conjugate image plane to the sample to generate multiple diffraction orders containing information about the electric-field at the object. An asymmetric mask (M) is used to physically completely block the $-1^{st}$ diffraction order from DG while passing the $+1^{st}$ diffraction order. Note that in the case SI, each diffraction order from the DG contains the $0^{th}$ and $\pm 1^{st}$ diffraction orders from the SLM. To generate a uniform wavefront reference beam necessary for off-axis holography, the mask also applies a pinhole spatial filter to the $0^{th}$ diffraction order from the SLM contained in the $0^{th}$ diffraction order from the DG. Due to the broadband spectrum of the NKT-illumination source, this diffraction order is the only undispersed component suitable for spatial filtering.

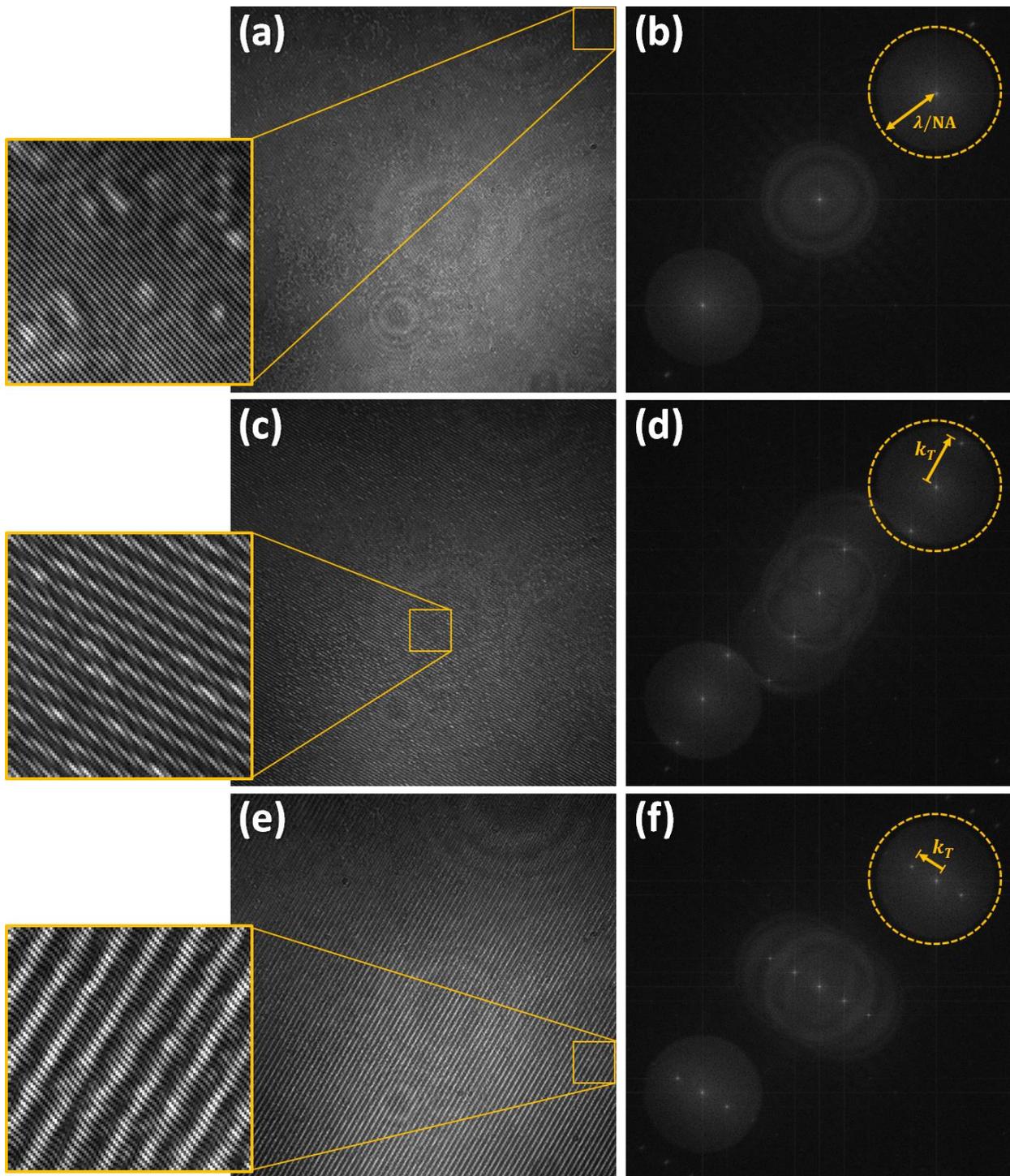

**Supplementary Figure 5.** Examples of (a,c,e) raw acquisitions are shown alongside associated (b,d,ef) Fourier distributions. As expected from off-axis holography techniques, the complex electric-field at the sample plane can be isolated from the autocorrelation and conjugate terms in Fourier space. Raw acquisition with associated Fourier distributions are shown for (a,b) widefield imaging, (c,d) SI imaging with sinusoidal spatial frequency and angle set at $|\mathbf{k}_T| = \text{NA}/\lambda$ and $\theta = \pi/3$, respectively, and (e,f) SI imaging with sinusoidal spatial frequency and angle set at $|\mathbf{k}_T| = \text{NA}/2\lambda$ and $\theta = 5\pi/6$, respectively.

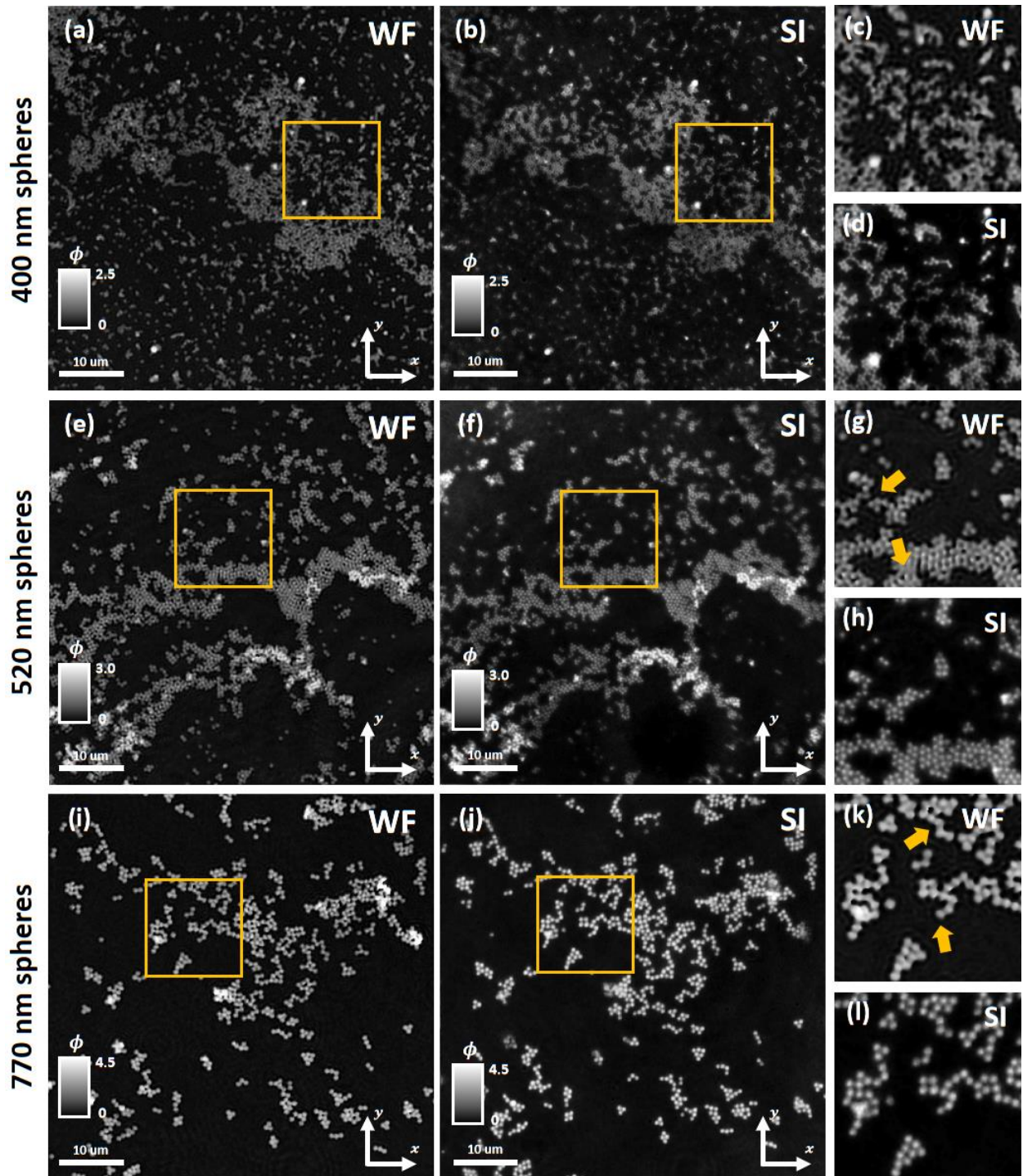

**Supplementary Figure 6.** (a,e,j) WF QP is compared with (b,f,j) SI QP when taking optical image sections through 400nm, 520nm, and 770nm diameter polystyrene spheres, respectively. Lateral sub-diffraction imaging capabilities are clearly demonstrated when comparing the WF and SI zooms (c and d, respectively) of the 400 nm beads. As demonstrated in the WF and SI zooms of the 520nm spheres (g and h, respectively), SI's tomographic capability is also important to reject any out-of-focus diffraction artifacts that can otherwise occlude sample features, even if such features are within the diffraction limit. Such artifacts may affect even features well within the diffraction limit, as shown by the diffraction rings and de-focused edges in the WF and SI zooms of the 770nm spheres (k and l, respectively). Examples of such artifacts affecting diffraction-limited visualization are indicated with yellow arrows in (g and k).

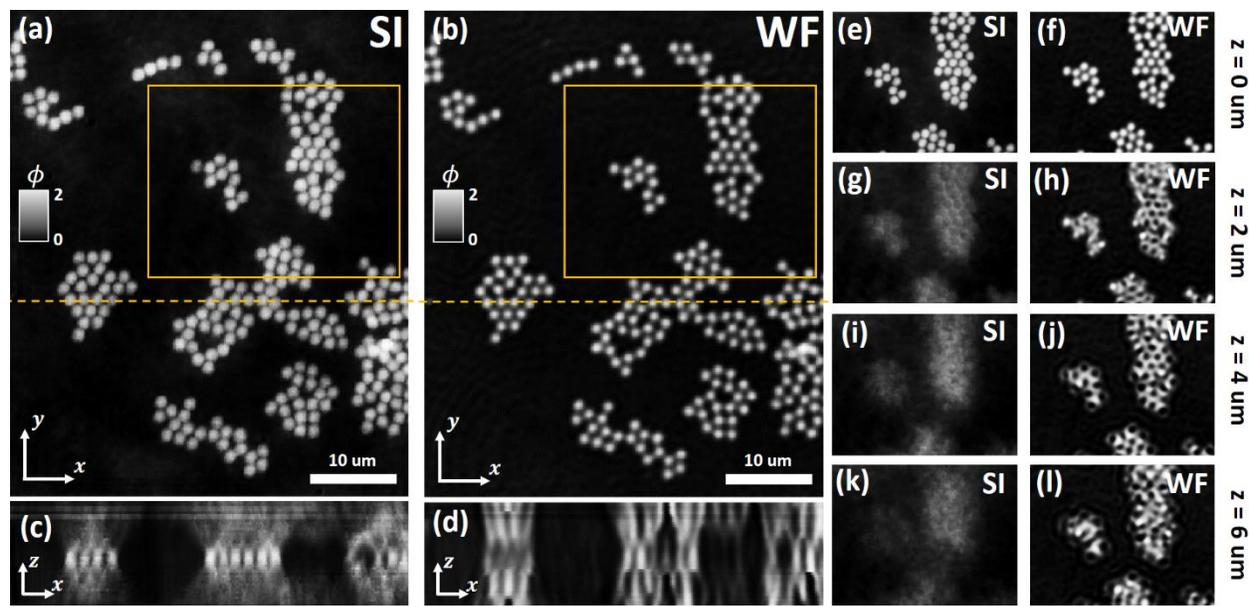

**Supplementary Figure 7.** 1.6um polystyrene microspheres are immersed in a refractive index medium oil of n = 1.51 and imaged through the 1.3 NA microscope objective. Sample features are well within the diffraction limit, and individual microspheres are clearly resolvable in both the (a) SI and (b) WF QP images. However, axial cuts through the location marked in dashed yellow in (a,b) clearly show that (c) SI QP imaging allows depth localization of the microspheres while (d) conventional WF QP imaging allows no such depth localization, even though the microspheres are clearly laterally resolvable. This is a direct result of the tomographic capability that SI QP enables over WF QP, as presented in Figure 2. To emphasize this point, we show zooms (from regions outlined in yellow in (a,b)) of defocused sample planes at 2um increments through the (e,g,i,k) SI and (f,h,j,l) WF QP volumes. In SI QP volume, the QP signal shows clear attenuation with increasing defocus, which is indicative of its depth localization. No such attenuation is present in the WF QP volume, where the QP signal from the microspheres instead diffracts out into the defocused planes.

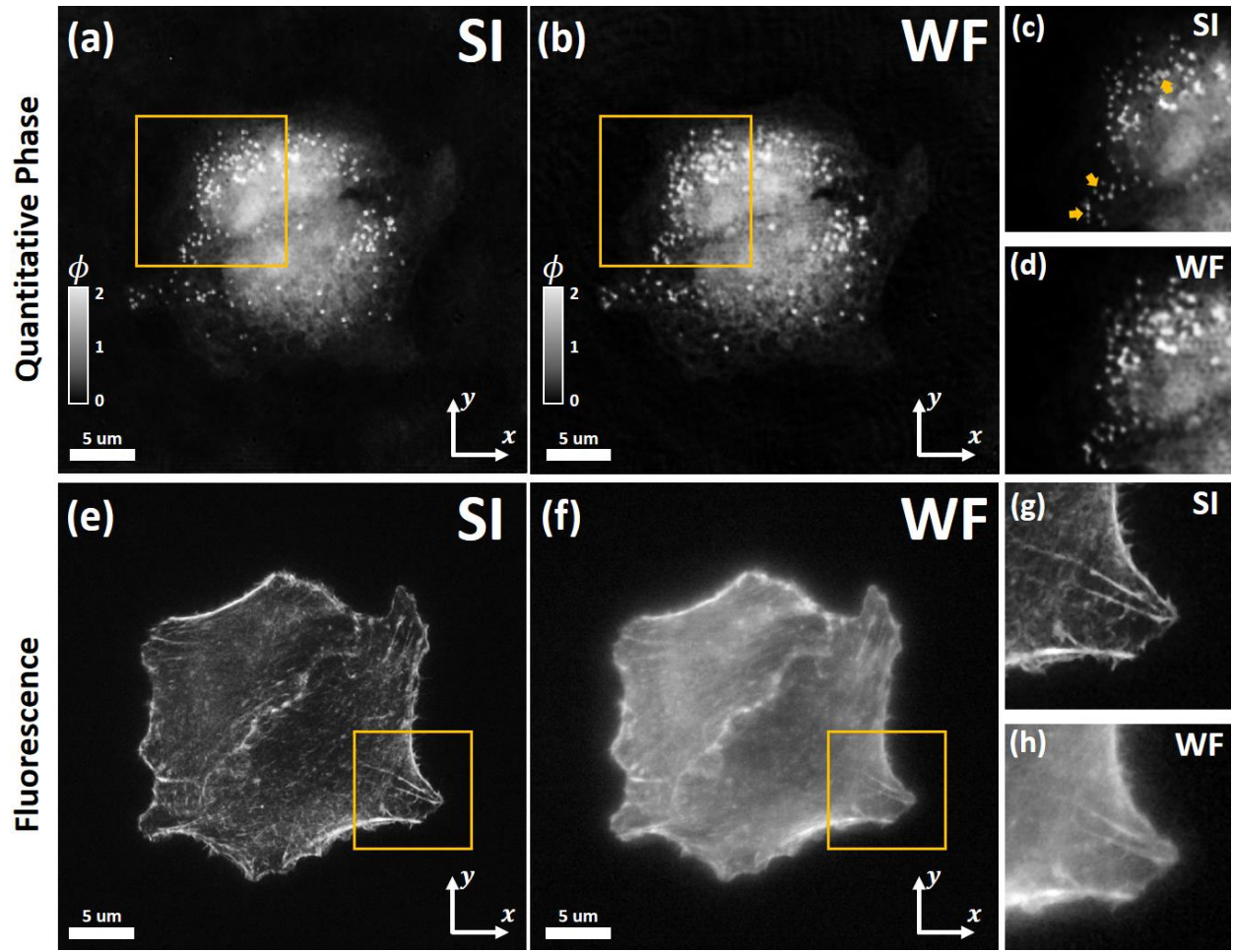

**Supplementary Figure 8.** Visualization capabilities of QP and fluorescent imaging are compared between the (a,e) SI and (b,f) conventional WF techniques when imaging two adjacent A549 cells. Lateral resolution gain is clearly evident when comparing the zooms of the (c,g) QP and fluorescent SI optical sections to their (d,h) WF counterparts, respectively. Enhanced SI QP resolution shows clear elucidation of high phase-delay structures in close proximity, indicated by yellow arrows in (c), that are not separable via WF imaging. In the SI enhanced fluorescence case, axial rejection of out of focus fluorescent portions reduces fluorescent haze, which otherwise obscures the F-actin branching at the junction between the two cells (which can also be visualized in the QP images).

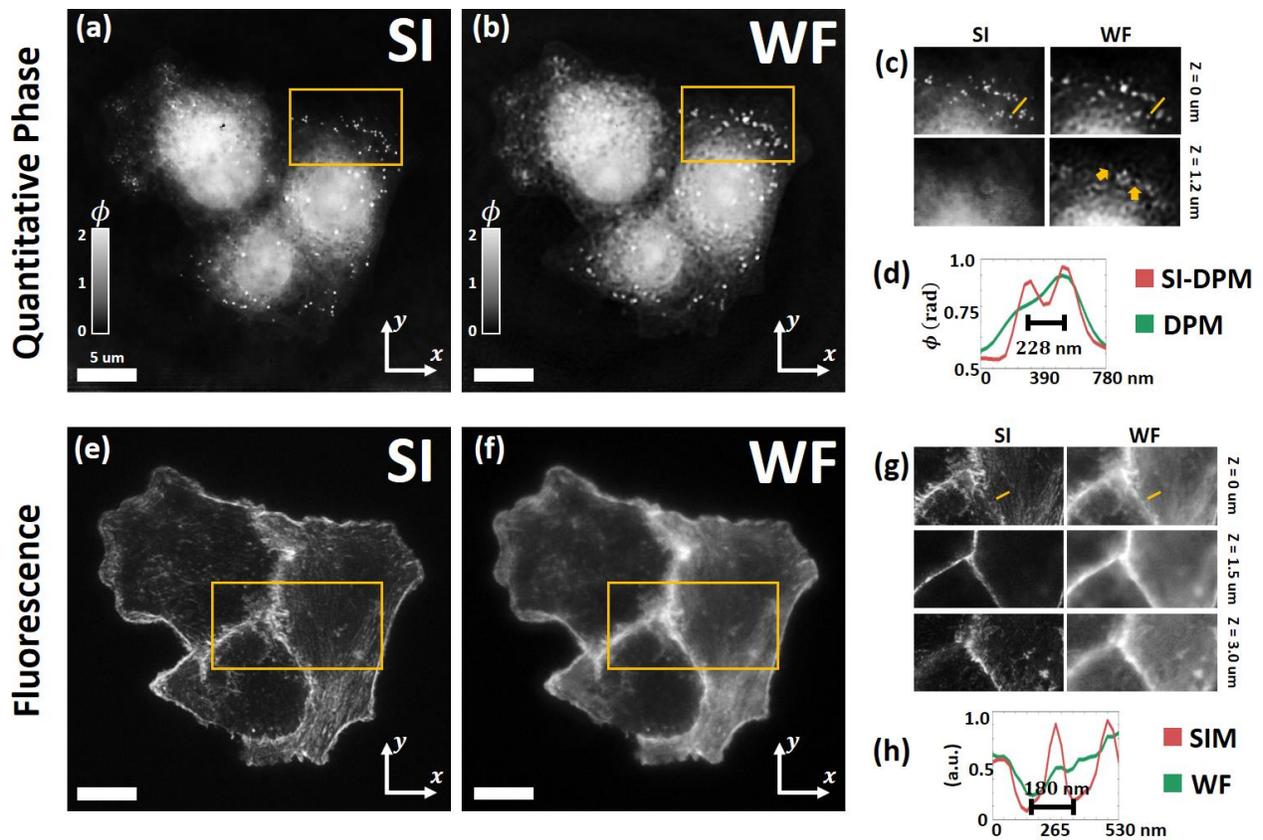

**Supplementary Figure 9.** Example is shown of SI 3D sub-diffraction resolving both fluorescence and quantitative-phase (QP) for an A549 cell cluster. (a,b) QP is shown with and without SI enhancement, respectively. (c) Zooms of selected regions from the SI-enhanced and normal widefield (WF) QP images are compared when sample is defocused by 1.2 um. Resolution increase is apparent in the in-focus QP zooms, and a QP profile plot resolves structures 228 nm apart. Comparison of QP images after defocus shows that, without SI enhancement, high-QP structure diffracts onto the defocused image plane. (d,e) The same system also performs 3D SI fluorescent super-resolution, as is demonstrated when comparing super-resolved and diffraction-limited WF fluorescent imaging of F-actin. (f) Zoom of selected region is shown undergoing defocus to 1.5 um and 3.0 um. Confirming what has been previously demonstrated[30], SI fluorescent super-resolution demonstrates out-of-focus rejection via filling of the missing cone. Intensity profile crosscut of a f-Actin filament shows 180 nm width between signal troughs.